\documentclass[11pt]{article}
\usepackage{graphicx}
\usepackage[small]{nicecite}
\usepackage{bm}
 \newcommand{\tn}{\tilde{n}}
 \newcommand{\td}{\tilde{\delta}}
\title{Phonon impact on the coherent
control of quantum states in semiconductor quantum dots}
\author{Anna Grodecka$^{1}$, Lucjan Jacak$^{1}$,\\
Pawe{\l} Machnikowski$^{1,2,}$\footnote{Corresponding author;
e-mail address: Pawel.Machnikowski@pwr.wroc.pl}, and Katarzyna
Roszak$^{1}$ \\
\small $^{1}$Institute of Physics, Wroc{\l}aw
University of Technology,\\ \small 50-370 Wroc{\l}aw, Poland\\
\small $^{2}$Institut f\"ur Festk\"orpertheorie, Westf{\"a}lische
Wilhelms-Universit\"at,\\ \small 48149 M\"unster, Germany}

\date{}

\begin{document}
\maketitle

\begin{abstract}
This chapter is devoted to the recent theoretical results on the
optical quantum control over charges confined in quantum dots
under influence of phonons. We show that lattice relaxation
processes lead to decoherence of the confined carrier states. The
theoretical approach leading to a uniform, compact description of
the phonon impact on carrier dynamics, perturbative in phonon
couplings but applicable to arbitrary unperturbed evolution, is
described in detail. Next, some applications are presented: phonon
damping of Rabi oscillations in quantum dots and phonon-induced
error of a single-qubit gate for an excitonic quantum dot qubit as
well as for a semiconductor quantum dot spin qubit operated via a
STIRAP transfer.
\end{abstract}

\section{Introduction}

With the state-of-the-art experimental techniques it is possible
to control the quantum state of charge carriers in a quantum dot
(QD). Many effects known from quantum optics of natural atoms have
been demonstrated in these man-made systems: controlled coherent
dynamics in these structures has been induced \cite{bonadeo98},
Rabi oscillations have been observed
\cite{stievater01,kamada01,zrenner02,borri02a,htoon02}) and
entanglement between states of interacting dots \cite{bayer01} has
been demonstrated.

However, unlike their natural counterparts, the artificial atoms
are solid-state structures, embedded in the surrounding
macroscopic crystal. Therefore, even in top-quality samples, some
perturbing interaction effects are inevitable. The mutual
influence of lattice deformation (phonons) and charge
distributions is one of these inherent effects. There are three
major mechanisms of carrier-phonon interaction \cite{mahan00}: (1)
Coulomb interaction with the lattice polarization induced by the
relative shift of the positive and negative sub-lattices of the
polar compound, described upon quantization by longitudinal
optical (LO) phonons; (2) deformation potential coupling
describing the band shifts due to lattice deformation, i.e. mainly
longitudinal acoustical (LA) phonons; (3) Coulomb interaction with
piezoelectric field generated by crystal deformation (LA and
transversal acoustical, TA, phonons). The lattermost effect is
weak for globally charge-neutral excitations (e.g. excitons) in
some systems, like InAs/GaAs, but may be of more importance for
uncompensated charge distributions (e.g. excess electrons) or for
the properties e.g. of GaN dots \cite{hohenester00d,derinaldis02},
where charges are separated by large built-in fields.

The coupling to the lattice degrees of freedom manifests itself in
many ways in the spectroscopic properties of QDs. Resonant
interaction with LO phonons leads to the extremely pronounced
spectrum reconstruction (formation of resonant polarons)
\cite{hameau99,hameau02,verzelen02a,jacak03a}, acoustic and
optical phonons provide a relaxation channel for carriers
\cite{heitz97,heitz01,ignatiev01} with an important role played by
phonon anharmonicity \cite{li99,verzelen00,jacak02a,verzelen02b},
phonon replicas and phonon-assisted transitions are magnified due
to resonant interaction with quantized carrier states
\cite{heitz99,heitz00,lemaitre01,findeis00,fomin00,jacak02f}.

The perturbing effect of lattice modes has also been observed
experimentally as a fast ($\sim 1$ ps) partial decay of coherent
optical polarization induced by an ultra-fast laser pulse
\cite{borri01,borri02a}. The theoretical analysis both in the
linear regime \cite{krummheuer02,jacak03b} and in the nonlinear
case \cite{vagov02a,vagov03} shows that this decay should be
viewed as a trace of coherent lattice dynamics (due to lattice
inertia) rather than as an effect of typical noise. The agreement
between the theoretical modeling of carrier-phonon kinetics
\cite{vagov03} and the experimental results \cite{borri01} shows
that phonon-related effects play the dominant role in the kinetics
of a confined system on picosecond timescales. Phonon effects are
also likely to play the leading role in the damping of Rabi
oscillations induced optically in a QD
\cite{forstner03,machnikowski03a,machnikowski04a} (although in the
current experiments other effects are also important). Due to
strong reservoir memory on timescales relevant for these
processes, they cannot be fully understood within the Markovian
approximations. In particular, the idea of a ``decoherence time'',
with which the control dynamics competes, is misleading
\cite{alicki02a}.

Apart from the general scientific interest, a strong motivation
for studying the phonon processes taking place in semiconductor
nanostructures comes from the recent proposals for defining a
quantum bit (qubit, the basic unit of quantum information) in
terms of orbital (charge) degrees of freedom of an exciton
confined in a quantum dot \cite{biolatti00} or in terms of spin
states controlled via conversion to orbital degrees of freedom
\cite{imamoglu99,pazy03a,troiani03,calarco03,feng03,chen04}. The
recent experimental demonstration of a quantum logic gate
operation with charge degrees of freedom in a QD \cite{li03} has
strengthened this motivation even more. Understanding
phonon-induced decoherence processes is essential for practical
implementation of these novel ideas and currently seems to be one
of the most topical issues in the field.

In order to study the phonon effects for the general
coherent-control and quantum information processing schemes, a
theoretical method going beyond the perturbative treatment of the
external driving is needed. In this chapter, we present one of
such methods, treating the Coulomb interactions between the
confined carriers and the coupling to the driving field exactly
(non-perturbatively), while the phonon coupling is included as a
perturbation. Although the range of validity of such a strictly
perturbative treatment may be narrower than that of more
sophisticated methods (e.g. the cumulant expansion technique
\cite{forstner03}), the advantage of the proposed approach is that
it yields closed formulas leading to a clear physical
interpretation of the phonon decoherence effects.

The chapter is organized as follows: The next section describes
the system under study, introduces its model and discusses the
unperturbed evolution under special conditions. In the section
\ref{sec:coupling} we describe the various carrier--phonon
coupling mechanisms for single carriers and for excitons. The
general idea of decoherence due to lattice relaxation is
introduced in the section \ref{sec:dress}. Next, in the section
\ref{sec:method} we develop the theoretical treatment for
analyzing the phonon-related decoherence for an arbitrary system
evolution. This is then applied to the description of phonon
damping of Rabi oscillations (section \ref{sec:rabi}),
optimization of control for an excitonic QD qubit (section
\ref{sec:opty}) and fidelity of a spin qubit operated via STIRAP
transition (section \ref{sec:stirap}). The final section concludes
the chapter.

\section{The system and the model}
\label{sec:model}

\paragraph{The Hamiltonian of the system.}
We will consider a system of charges confined in a QD, coupled to
coherent electromagnetic field and interacting with phonons of the
surrounding crystal medium. The Hamiltonian for the system is
\begin{equation}\label{ham-gener}
    H = H_{\mathrm{C}}+H_{\mathrm{ph}}+H_{\mathrm{int}}.
\end{equation}
The first term $H_{\mathrm{C}}$ describes the carrier subsystem
together with the external driving field. Throughout this chapter,
we will assume that this driving field is strong enough and its
quantum fluctuations may be neglected, so that the field may be
modeled classically. Usually, for a confined system only the
lowest part of the discrete spectrum is relevant for reasonable
pulse durations and intensities, so it is often convenient to
assume that the total wavefunctions of the interacting carrier
system are known (e.g. from numerical diagonalization
\cite{jacak03b}) and to express the Hamiltonian in the basis of
these eigenstates. Thus, unless explicitly stated otherwise, we
will denote the crystal vacuum by $|0\rangle$ and the confined
exciton states by $|n\rangle$, $n\ge 1$.

Numerical calculations for the harmonic confinement model
\cite{jacak03b} show that the lowest exciton states may be well
approximated by products of the electron ground-state harmonic
oscillator wavefunction and a hole wavefunction corresponding to
one of the low harmonic oscillator levels. This is due to the
large hole effective mass compared to that of the electron. For
the same reason, even if the geometrical confinement is the same
for both particles, the Coulomb interaction shrinks the hole
wavefunction while the electron wavefunction is only slightly
modified (the electron excitation energy is much larger than
Coulomb energy in a typical self-assembled structure). The
separation between the ground state and the lowest dark excited
state (hole excitation) is usually of several meV, while the
closest bright state (the lowest one with the electron in an
excited state) lies approximately at 70 meV.

Also, for the structure of excitonic levels as discussed above,
the single-exciton assumption is reasonable under excitation with
polarized light. Even though excited single-exciton states may be
close to the ground state, they are formed by exciting the hole
(which is much heavier than the electron), while the electron
remains roughly in its ground single-particle state. Creation of a
bi-exciton requires much higher energy, sufficient to promote both
carriers to excited states.

The second term in (\ref{ham-gener}) is
\begin{equation}\label{H-ph}
    H_{\mathrm{ph}}
     =\sum_{\bm{k}}\omega_{\bm{k}}b_{\bm{k}}^{\dagger}b_{\bm{k}}
\end{equation}
and describes the energies of phonons,
$b_{\bm{k}}^{\dag},b_{\bm{k}}$ being the phonon creation and
annihilation operators (branch index will always be included into
$\bm{k}$, unless explicitly written). Since only long-wavelength
phonons are effectively coupled to carriers confined in a QD (see
discussion in the next section) we will here always assume linear
and isotropic dispersion for acoustic phonons, so that
$\omega_{s}(\bm{k})=c_{s}k$, where $c_{s}$ is the speed of sound
for the branch $s$ ($s=\mathrm{l}$ for LA, $s=\mathrm{t}$ for TA).
Also, because the dispersion of the LO phonons is weak around the
zone center, the LO phonons ($s=\mathrm{o}$) will be assumed
dispersionless, $\omega_{\mathrm{o}}=\Omega$. We will neglect any
anharmonicity effects and assume the free harmonic evolution of
the lattice subsystem in absence of the carrier-phonon coupling.
Only the bulk phonon modes will be included. Although a quantum
dot implies certain electrostatic and mechanical discontinuity of
the system and the role of confined phonons may be discussed
\cite{li99,vasilevskiy04}, invoking bulk phonons is usually
sufficient for explanation of experimental features, like e.g.
polaron resonances \cite{hameau99,hameau02}.

The last term,
\begin{equation}\label{H-int}
    H_{\mathrm{int}}=\sum_{\bm{k};n,n'\ge 1}|n\rangle\!\langle n'|
    F_{nn'}(\bm{k})\left(b_{\bm{k}}^{\dagger}+b_{-\bm{k}}\right),
    \;\;\; F_{nn'}(\bm{k})=F_{nn'}^{*}(-\bm{k}),
\end{equation}
is the exciton-phonon interaction expressed in terms of the
discrete confined carrier states $|n\rangle$. Although inter-band
phonon terms may appear e.g. as a consequence of strain-dependent
inter-band couplings, in the following sections we deal only with
intraband phonon effects, so that the interaction Hamiltonian
conserves the number of quasiparticles of each kind.

The phonon wavevectors are restricted to the first Brillouin zone.
However, modulations of the band structure with spatial periods
much smaller than the size of the carrier wavefunction cannot
influence the carrier energy. In the harmonic oscillator
approximation for carrier confinement, this leads to exponential
cut-off of carrier-phonon interaction (explicit formulas are given
in the section \ref{sec:coupling} below) at the frequency
$\omega_{0}\simeq c/l$, where $c$ is the sound speed, $l$ is the
carrier confinement size. Thus, the exciton is effectively coupled
only to the long-wavelength part of the phonon modes. On the other
hand, vanishing of the coupling in the limit of $\bm{k}=0$
reflects the insensitivity of the system properties to shifting
the lattice as a whole. For gapless bosons with linear dispersion,
the characteristic frequency determines the reservoir memory times
and sets up the timescale of non-Markovian effects, related to
``dressing'' of the localized carriers with coherent lattice
deformation field (see section \ref{sec:dress}).

\paragraph{Unperturbed resonantly driven evolution}
Assuming that the laser field couples resonantly only to one
transition (say, from the ``empty dot'' state $|0\rangle$ to the
ground exciton state $|1\rangle$) the Hamiltonian for the carrier
subsystem may be written
\begin{equation}\label{ham-C-rezon}
    H_{\mathrm{C}0}=\sum_{n}E_{n}|n\rangle\!\langle n|
        +\frac{1}{2}f(t)\left(e^{i\omega t}|1\rangle\!\langle 0|
            +e^{-i\omega t}|0\rangle\!\langle 1|\right)
\end{equation}
(the energies $E_{n}$ are defined with respect to the ``empty
dot'' state).

Using the canonical transformation to the ``rotating frame'',
defined by the unitary operator
\begin{displaymath}
    U_{\mathrm{rw}}
      =\exp\left[
        -i\omega \sum_{n\ge 1}|n\rangle\!\langle n| t\right],\;\;\;
    U_{\mathrm{rw}}|n\rangle = e^{-i\omega t}|n\rangle,\;\;n\ge 1,
\end{displaymath}
we obtain from (\ref{ham-C-rezon})
\begin{equation}
  H_{\mathrm{C}}  =
    U^{\dagger}_{\mathrm{rw}}H_{\mathrm{C}0}U_{\mathrm{rw}}
        -\sum_{n\ge 1}|n\rangle\!\langle n|
 = -\sum_{n\ge 1}\Delta_{n} |n\rangle\!\langle n|
    +\frac{1}{2}f(t) \left( |1\rangle\!\langle 0|
            +|0\rangle\!\langle 1| \right)
\label{ham-rotframe}
\end{equation}
where $\Delta_{n}=\omega-E_{n}$. In the special case of strictly
resonant coupling to the ground-state excitonic transition,
$\omega=E_{1}$, so that the detuning from the ground-state
transition vanishes, $\Delta_{1}=0$, and
$-\Delta_{n}=\epsilon_{n}$ where $\epsilon_{n}$ are the intra-band
excitation energies for a single exciton.

Let us denote the evolution operator for the exciton--light system
(generated by $H_{\mathrm{C}}$, without phonon interactions) by
$U_{0}(t,s)$, where $s,t$ are the initial and final times,
respectively. In the special case of resonant coupling only
between $|0\rangle$ and $|1\rangle$ state the evolution operator
may be found explicitly (note that in this resonant case
$H_{\mathrm{C}}$ commutes with itself at different times). The
result is
\begin{equation}\label{evol-rezon}
    U_{\mathrm{C}}(t,s)=
        \cos\frac{\Phi(t)}{2}(|0\rangle\!\langle0|
                       +|1\rangle\!\langle 1|)
            -i\sin\frac{\Phi(t)}{2}(|0\rangle\!\langle 1|
                       +|1\rangle\!\langle 0|)
        +\sum_{n>1}|n\rangle\!\langle n|e^{-i\epsilon_{n}(t-s)},
\end{equation}
where
\begin{displaymath}
    \Phi(t)=\int_{s}^{t}d\tau f(\tau)
\end{displaymath}
is the rotation angle on the Bloch sphere up to time $t$.

\section{Interactions of confined carriers with phonons}
\label{sec:coupling}

For completeness, in this section we summarize the derivation of
the coupling constants between the bulk phonon modes and the
confined carriers in a polar and piezoelectric semiconductor
\cite{mahan00,haken76,mahan72}.

\paragraph{The deformation potential.}
Any crystal deformation leads to shifts of the conduction (c) and
valence (v) bands which are, to the leading order, proportional to
the relative volume change. The corresponding contribution to the
energy of electrons (e) and holes (h) in the long-wavelength limit
is, therefore,
\begin{displaymath}
    H^{(\mathrm{DP})}_{\mathrm{e/h}} \equiv \pm\Delta E_{\mathrm{c/v}}
    =\mp\sigma_{\mathrm{e/h}}\frac{\delta V}{V},
\end{displaymath}
where $\sigma_{\mathrm{e/h}}$ are the deformation potential
constants for electrons and holes and $V$ is the unit cell volume.
Using the strain tensor $\hat{\sigma}$,
\begin{displaymath}
    \sigma _{ij}=\frac{1}{2}\left(
\frac{\partial u_i}{\partial r_j} +\frac{\partial u_j}{\partial
r_i}\right),
\end{displaymath}
one may write
\begin{displaymath}
    H^{(\mathrm{DP})}_{\mathrm{e/h}}
        =\mp\sigma_{\mathrm{e/h}} \mathrm{Tr}\hat{\sigma}
        =\mp\sigma_{\mathrm{e/h}} \nabla\cdot\bm{u}(\bm{r}),
\end{displaymath}
where $\bm{u}(\bm{r})$ is the local displacement field. The
displacement is quantized in terms of phonons,
\begin{equation}\label{phonon}
    \bm{u}(\bm{r})=i\frac{1}{\sqrt{N}}\sum_{\bm{k}}
        \sqrt{\frac{\hbar}{2\rho V \omega(\bm{k})}}
        \hat{\bm{e}}_{\bm{k}}\left(
            b_{\bm{k}}+b^{\dag}_{-\bm{k}}
        \right)e^{i\bm{k}\cdot\bm{r}},
\end{equation}
where $\omega(\bm{k})$ is the frequency for the wavevector
$\bm{k}$, $\hat{\bm{e}}_{\bm{k},s}=-\hat{\bm{e}}_{-\bm{k},s}$ is
the corresponding real unit polarization vector, and $\rho$ is the
crystal density. Only the longitudinal branch contributes to
$\nabla\cdot\bm{u}$ in (\ref{phonon}) and the final interaction
Hamiltonian in the coordinate representation for carriers is
\begin{equation}\label{int-DP-r}
    H^{(\mathrm{DP})}_{\mathrm{e/h}}
        =\pm\sigma_{\mathrm{e/h}}\frac{1}{\sqrt{N}}\sum_{\bm{k}}
        \sqrt{\frac{\hbar k}{2\rho V \omega_{\mathrm{l}}(\bm{k})}}
        \left( b_{\bm{k},\mathrm{l}}+b^{\dag}_{-\bm{k},\mathrm{l}}
        \right)e^{i\bm{k}\cdot\bm{r}}.
\end{equation}

In the second quantization representation with respect to the
carrier states this reads
\begin{eqnarray}
    H_{\mathrm{e/h}}^{(\mathrm{DP})}
  & = & \sum_{nn'}\langle n|V(\bm{r})|n'\rangle a_{n}^{\dag}a_{n'}
    = \sum_{nn'}
    \int_{-\infty}^{\infty} d^3r\psi _n^* (\bm{r}) V(\bm{r})
    \psi_{n'}(\bm{r}) a_{n}^{\dag}a_{n'} \nonumber \\
  & = & \frac{1}{\sqrt{N}}\sum_{\bm{k}nn'}a_{n}^{\dag}a_{n'}
  f_{\mathrm{e/h},nn'}^{(\mathrm{DP})}(\bm{k})
\left(b_{\bm{k},\mathrm{l}}+b_{-\bm{k},\mathrm{l}}^{\dag}\right),
\label{int-DP}
\end{eqnarray}
where
\begin{displaymath}
    f_{\mathrm{e/h},nn'}^{(\mathrm{DP})} (\bm{k})
    =\pm\sigma_{\mathrm{e/h}}\sqrt{\frac{\hbar k}{2\rho V c_{\mathrm{l}}}}
    \mathcal{F}_{nn'} (\bm{k}),
\end{displaymath}
with the formfactor
\begin{equation}\label{formfactor}
    \mathcal{F}_{nn'}(\bm{k})
       =\int_{-\infty}^{\infty} d^3\bm{r}\psi _n^*
        (\bm{r})e^{i\bm{k}\cdot\mathrm{r}}\psi_{n'}(\bm{r})
       =\mathcal{F}_{n'n}^{*}(-\bm{k}).
\end{equation}
The interaction Hamiltonian (\ref{int-DP}) conforms with the
general form assumed in (\ref{H-int}).

\paragraph{General properties of the formfactors.}
While the common coefficient of the coupling Hamiltonian contains
the fundamental and material-dependent constants and reflects the
general electrical and mechanical properties of the semiconductor
system, the formfactor (\ref{formfactor}) contains the information
about the geometry of the confinement and the resulting properties
of wavefunctions. In this sence, it is the ``engineerable'' part
of the carrier-phonon coupling.

From orthogonality of single-particle states one has immediately
$\mathcal{F}_{nn'}(0)=\delta_{nn'}$. If the wavefunctions are
localized at a length $l$, then the extent of the formfactor is
$\sim 1/l$. Thus, for carrier states localized in a QD over many
lattice sites and smooth within this range, the functions
$\mathcal{F}_{nn'}(\bm{k})$ will be localized in the
$\bm{k}$--space very close to the center of the Brillouin zone.

As an example, let us consider the ground state of the harmonic
oscillator potential,
\begin{equation}
    \psi(\bm{r})=\frac{1}{\pi^{3/4}l_{z}l_{\bot}}
      \exp\left[-\frac{1}{2}\left(\frac{r_{\bot}}{l_{\bot}}\right)^{2}
        -\frac{1}{2}\left( \frac{z}{l_{z}}\right)^{2}\right],
\label{ground-state}
\end{equation}
where $r_{\bot}$ is the position component in the $xy$ plane and
$l_{\bot},l_{z}$ are the localization widths in-plane and in the
growth ($z$) direction. The corresponding formfactor is then
easily found to be
\begin{equation}\label{formf-expli}
    \mathcal{F}(\bm{k})=\exp\left[
        -\left( \frac{k_{\bot}l_{\bot}}{2}\right)^{2}
        -\frac{1}{2}\left( \frac{k_{z}l_{z}}{2}\right)^{2}\right].
\end{equation}

\paragraph{Piezoelectric interaction.}
A propagating phonon wave in a piezoelectric medium induces a
polarization field which affects the carriers by means of the
Coulomb interaction. If the crystal deformation is described by
the strain tensor $\hat{\sigma}$ then the piezoelectric
polarization is $\bm{P}=\hat{d}\hat{\sigma}$, where $\hat{d}$ is
the piezoelectric tensor.

From the Maxwell equation for plane-wave fields (in absence of
external charges and currents),
\begin{displaymath}
    \begin{array}{ll}
    i\bm{k} \cdot \bm{E}=-\frac{1}{\varepsilon _0}
        i\bm{k}\cdot \bm{P},
 &  i\bm{k}\times \bm{E} =i\omega \bm{B}, \\
    i\bm{k} \cdot \bm{B}=0,
 &  c^2 i\bm{k} \times \bm{B}=-i\omega\bm{E}-
        \frac{1}{\varepsilon _0}i\omega \bm{P},
    \end{array}
\end{displaymath}
one has
\begin{displaymath}
    c^2 \frac{1}{\omega}
        [\bm{k} (\bm{k} \cdot \bm{E})-k^2\bm{E}]
    =-\omega[
\bm{E}+ \frac{1}{\varepsilon _0} \varepsilon _0\kappa \bm{E}+
\frac{1}{\varepsilon _0}\hat{d} \hat{\sigma}],
\end{displaymath}
where we have used the relation $\bm{P}=\varepsilon_{0}\kappa
\bm{E}+\hat{d}\hat{\sigma}$.

For the transversal component we get
\begin{displaymath}
    c^2 \frac{1}{\omega}k^2 \bm{E}_{\bot }
        =-\omega\left[ \bm{E}_{\bot }(1+\kappa )
        +\frac{1}{\varepsilon _0}(\hat{d}\hat{\sigma})_\bot
        \right],
\end{displaymath}
hence,
\begin{displaymath}
    E_{\bot}\sim\left(\frac{\omega}{c k}\right)^{2}.
\end{displaymath}
For strain fields associated with phonon propagation, $\omega$ and
$k$ are phonon frequency and wavevector and $\omega/k=c_{s}\ll c$
so that the transversal component vanishes. Thus, the
piezoelectric field accompanying a phonon wave is purely
longitudinal and one may introduce the appropriate potential. One
has explicitly for the longitudinal component
\begin{displaymath}
    \pm eE_{\parallel }=
    \mp\frac{e}{\varepsilon _0} \frac{(\hat{d}
        \hat{\sigma})_\parallel }{1+\kappa }=-\nabla V(\bm{r}),
\end{displaymath}
where
\begin{displaymath}
    V(\bm{r})=-i\frac{\pm e}{k} \frac{(\hat{d}
\hat{\sigma})_\parallel }{\varepsilon_0 \varepsilon_1 }.
\end{displaymath}
Using (\ref{phonon}), one finds the strain tensor
\begin{displaymath}
    \sigma _{ij}=-\frac{1}{2}\frac{1}{\sqrt{N}}\sum_{s,\bm{k}}
    \sqrt{\frac{\hbar}{2M\omega_{s,\bm{k}}}}(
    \hat{b}_{s,\bm{k}}+
    \hat{b}^+_{s,-\bm{k}})\left[(\hat{e}_{s,\bm{k}}
    )_i k_j+(\hat{e}_{s,\bm{k}})_j
    k_i\right]e^{i\bm{k}\cdot\bm{r}}.
\end{displaymath}
In the zincblende structure, one has
\begin{displaymath}
    d_{xyz}=d_{yzx}=d_{zxy}=d,\;\;\;d_{ijk}=d_{ikj},
\end{displaymath}
and the other components vanish. Hence
\begin{displaymath}
    (\hat{d} \hat{\sigma})_\parallel
    =2\frac{d}{k}(k_x\sigma_{yz}+k_y\sigma_{zx}+k_z\sigma_{xy})
\end{displaymath}
and
\begin{equation}\label{int-PE-r}
    H^{(\mathrm{PE})}_{\mathrm{e/h}}
        = \mp i\frac{1}{\sqrt{N}}\sum_{s,\bm{k}}
        \sqrt{\frac{\hbar}{2M\omega_{s}(\bm{k})}}
        \frac{de}{\varepsilon_0\varepsilon_1} M_{s}(\hat{\bm{k}})
        (b_{\bm{k},s}+b^{\dag}_{-\bm{k},s})
        e^{i\bm{k}\cdot\bm{r}},
\end{equation}
where the polarization-dependent geometrical factor is
\begin{displaymath}
    M_{s}(\hat{\bm{k}})=2\left[
        \hat{k}_{x}\hat{k}_{y}(\hat{e}_{s,\bm{k}})_{z}
        +\hat{k}_{y}\hat{k}_{z}(\hat{e}_{s,\bm{k}})_{x}
        +\hat{k}_{z}\hat{k}_{x}(\hat{e}_{s,\bm{k}})_{y}
                    \right].
\end{displaymath}
For the choice of the phonon polarizations (l--longitudinal,
t1,t2--transversal)
\begin{eqnarray*}
    \hat{e}_{\mathrm{l},\bm{k}} &\equiv& \hat{\bm{k}}
        = (\cos\theta\cos\phi,\cos\theta\sin\phi,\sin\theta),\\
    \hat{e}_{\mathrm{t1},\bm{k}} & = & (-\sin\phi,\cos\phi,0),\\
    \hat{e}_{\mathrm{t2},\bm{k}} & = &
        (\sin\theta\cos\phi,\sin\theta\sin\phi,-\cos\theta),
\end{eqnarray*}
the functions $M_{s}$ are
\begin{eqnarray}
\label{Ml}
  M_{\mathrm{l}}(\theta,\phi)  &=&
    \frac{3}{2}\sin2\theta\cos\theta\sin2\phi, \\
\label{Mt1}
  M_{\mathrm{t1}}(\theta,\phi) &=& \sin2\theta\cos2\phi, \\
\label{Mt2}
  M_{\mathrm{t2}}(\theta,\phi) &=&
    (3\sin^{2}\theta-1)\cos\theta\sin 2\phi.
\end{eqnarray}
Finally, in the second quantization representation,
\begin{eqnarray}\label{int-PE}
    H^{(\mathrm{PE})}_{\mathrm{e/h}} & = & \sum_{nn'}a_{n}^{\dag}a_{n}
        \int_{-\infty}^{\infty}d^3\bm{r}
         \psi_{n}^{*} (\bm{r})V(\bm{r})\psi_{n'}(\bm{r})\\
        & = &\frac{1}{\sqrt{N}}\sum_{\bm{k},s,nn'}a_{n}^{\dag}a_{n'}
  f_{\mathrm{e/h},nn',s}^{(\mathrm{PE})}(\bm{k})
\left(b_{\bm{k},s}+b_{-\bm{k},s}^{\dag}\right),\nonumber
\end{eqnarray}
where
\begin{displaymath}
    f_{\mathrm{e/h},nn',s}^{(\mathrm{PE})} (\bm{k})
    =\mp i\sqrt{\frac{\hbar}{2\rho V\omega_{s}(\bm{k})}}
    \frac{de}{\varepsilon_{0}\varepsilon_{1}}M_{s}(\hat{\bm{k}})
    \mathcal{F}_{nn'} (\bm{k}),
\end{displaymath}
with the formfactor given by (\ref{formfactor}).

Note that the LA phonons are coupled to carriers both by the
deformation potential and by the piezoelectric coupling, while the
TA phonons only by the latter. The piezoelectric coupling results
directly from the Coulomb interaction and has exactly opposite
value for the electron and the hole. In contrast, there is in
general no fixed relation between the values of the deformation
potential constants.

\paragraph{Fr\"ohlich coupling}
An LO phonon propagating in a polar medium is accompanied by a
polarization field resulting from the relative shifts of the
positive and negative ions forming the crystal lattice. Similarly
as in the case of piezoelectric coupling, one shows that the
resulting electric field is longitudinal and may be associated
with a potential. Both the polarization and electric field, and
hence the potential, are proportional to the LO phonon
displacement. The derivation of the proportionality constant may
be found in \cite{mahan00,haken76}. The resulting interaction
energy for the charge carrier at point $\bm{r}$ is
\begin{equation}
H_{\mathrm{e/h}}^{(\mathrm{Fr})}=
\mp\frac{1}{\sqrt{N}}\sum_{\bm{k}}\frac{e}{k}
\sqrt{\frac{\hbar\Omega}%
{2v\epsilon_{0}\tilde{\epsilon}}} \left( b_{\mathrm{o},\bm{k}}
+b_{\mathrm{o},-\bm{k}}^{\dagger}\right) e^{i\bm{k}\cdot\bm{r}},
\label{int-Fr-r}
\end{equation}
where
$\tilde{\epsilon}=(1/\epsilon_{\infty}-1/\epsilon_{\mathrm{s}})^{-1}$
is the effective dielectric constant.

In the occupation number representation this reads
\begin{equation}\label{int-Fr}
    H^{(\mathrm{Fr})}_{\mathrm{e/h}}
        =\frac{1}{\sqrt{N}}\sum_{\bm{k}nn'}a_{n}^{\dag}a_{n'}
  f_{\mathrm{e/h},nn'}^{(\mathrm{PE})}(\bm{k})
\left(b_{\bm{k},l}+b_{-\bm{k},l}^{\dag}\right),
\end{equation}
where
\begin{displaymath}
    f^{(\mathrm{Fr})}_{\mathrm{e/h},nn'} =\mp\frac{e}{k}
    \sqrt{\frac{\hbar\Omega}{2v\epsilon_{0}\tilde{\epsilon}}}
    {\mathcal{F}}_{nn'}(\bm{k}).
\end{displaymath}

\paragraph{Phonon coupling for exciton states}
Above, we have derived the interaction Hamiltonian in the
single-particle basis. However, most of the following deals with
excitonic states, i.e. states of confined electron-hole pairs
interacting by Coulomb potentials.

Both carriers forming the exciton couple to phonons according to
(\ref{int-DP}), (\ref{int-PE}) and (\ref{int-Fr}). If the lowest
exciton states are assumed to be approximately of product form, as
discussed above, i.e.
\begin{displaymath}
    |n\rangle
        =a_{\mathrm{e},1}^{\dag}a_{\mathrm{h},n}^{\dag}|0\rangle,
\end{displaymath}
then one gets from (\ref{int-DP}) the following coupling constants
for the deformation potential interaction in the excitonic basis
\begin{eqnarray}\label{cpl-X-1}
  F^{(\mathrm{DP})}_{11}(\bm{k}) &=&
    \sqrt{\frac{\hbar k}{2\rho V c_{\mathrm{l}}}}\left(
    \sigma_{\mathrm{e}}\mathcal{F}^{(\mathrm{e})}_{11}(\bm{k})
    -\sigma_{\mathrm{h}}\mathcal{F}^{(\mathrm{h})}_{11}(\bm{k})
    \right), \\
\label{cpl-X-2}
  F^{(\mathrm{DP})}_{1l}(\bm{k}) &=&
    -\sqrt{\frac{\hbar k}{2\rho V c_{\mathrm{l}}}}
        \sigma_{\mathrm{h}}
        \mathcal{F}^{(\mathrm{\mathrm{h}})}_{1l}(\bm{k}),
            \;\;l>1.
\end{eqnarray}
It is essential to note that, due to different deformation
potential constants $\sigma_{\mathrm{e/h}}$, none of these
couplings vanishes even if the electron and hole wavefunctions are
the same, leading to identical single-particle formfactors.

For the Fr\"ohlich coupling to LO phonons one has
\begin{eqnarray}\label{cpl-X-3}
  F^{(\mathrm{Fr})}_{11}(\bm{k}) &=&
    -\frac{e}{k}
    \sqrt{\frac{\hbar\Omega}{2v\epsilon_{0}\tilde{\epsilon}}}
    \left( \mathcal{F}^{(\mathrm{e})}_{11}(\bm{k})
    -\mathcal{F}^{(\mathrm{h})}_{11}(\bm{k}) \right), \\
\label{cpl-X-4}
  F^{(\mathrm{Fr})}_{1l}(\bm{k}) &=&
    \frac{e}{k}
    \sqrt{\frac{\hbar\Omega}{2v\epsilon_{0}\tilde{\epsilon}}}\
        \mathcal{F}^{(\mathrm{\mathrm{h}})}_{1l}(\bm{k}),
            \;\;\;l>1.
\end{eqnarray}
The form of the coupling constants for the piezoelectric
interaction is analogous. In the case of Fr\"ohlich and
piezoelectric coupling, the ``diagonal'' coupling $F_{11}$
vanishes if the wavefunctions overlap exactly. This results from
the fact that both these interaction mechanisms are related
directly to the Coulomb interaction between the confined charge
distribution and the phonon-related polarization field. For
carriers localized in the same spatial volume but overlapping only
partly there will be still some cancelation effect. It is
remarkable, however, that even for perfect cancelation of the
diagonal coupling, the other contributions $F_{1n}$, $n>1$ are not
decreased. As we will see, this will lead to additional effects
for LO phonons, while in the case of piezoelectric coupling these
contributions are negligible due to large energy level spacing
compared to the acoustic phonon frequencies.

The spectral properties of the lattice are characterized by the
phonon spectral densities
\begin{equation}\label{spdens-expli}
    R_{nn',mm'}(\omega)=\frac{1}{\hbar^{2}}
    |n_{\mathrm{B}}(\omega)+1|\frac{1}{N}
    \sum_{\bm{k},s}F_{nn'}^{(s)}(\bm{k})F^{(s)*}_{m'm}(\bm{k})
    \left[\delta(\omega-\omega_{s}(\bm{k}))
    +\delta(\omega+\omega_{s}(\bm{k})) \right],
\end{equation}
where $n_{\mathrm{B}}(w)=-n_{\mathrm{B}}(-\omega)-1$ is the Bose
distribution function. These functions depend on the material
parameters and system geometry and fully characterize the
properties of the lattice subsystem at the level of perturbation
treatment discussed in this chapter. As will be seen later, they
are one of the two ``building blocks'' for the perturbative
calculation of the phonon effects on a quantum evolution. Here, we
will need only a subset of these functions, $R_{l}(\omega)\equiv
R_{1l,l1}(\omega)$.

Let us consider the functions $R_{l}^{(\mathrm{LA})}(\omega)$,
corresponding to  the LA phonons. Here the total coupling
$F_{1l}^{(\mathrm{LA})}$ contains both DP and PE contributions.
However, under reasonable symmetry assumptions the formfactors may
be chosen either real or purely imaginary [see e.g. the explicit
form for the harmonic confinement, Eq. (\ref{formf-expli})], while
the coefficients in (\ref{int-DP}) and (\ref{int-PE}) are purely
real and purely imaginary, respectively. Thus, the spectral
densities split into two independent contributions. Substituting
(\ref{cpl-X-1},\ref{cpl-X-2}) into (\ref{spdens-expli}) and
performing the summation over $\bm{k}$ in the usual continuum
limit
\begin{displaymath}
    \sum_{\bm{k}}\to \frac{NV}{(2\pi)^{2}}\int d^{3}\bm{k},
\end{displaymath}
one obtains
\begin{equation}\label{spdens-DP-onedot}
    R_{l}^{(\mathrm{DP})}(\omega)=R_{\mathrm{DP}}\omega^{3}
        [n_{\mathrm{B}}(\omega)+1] f_{l}(\omega),
\end{equation}
where
\begin{displaymath}
    R_{\mathrm{DP}}
     =\frac{(\sigma_{\mathrm{e}}-\sigma_{\mathrm{h}})^{2}}%
        {4\pi^{2}\hbar\rho c_{\mathrm{l}}^{5}}
\end{displaymath}
and $f_{l}(\omega)$ are certain functions that depend on the
wavefunction geometry, having the property $f_{1}(0)= 1$ and
$f_{l}(0)= 0$, $l>1$. A similar procedure leads to an analogous
result for the piezoelectric contribution from LA phonons.

In the case of dispersionless LO phonons
$\omega_{\mathrm{o}}(\bm{k})=\Omega$ and the spectral densities
are
\begin{eqnarray*}
    R^{(\mathrm{Fr})}_{1}(\omega) & = & R_{\mathrm{Fr}}
    |n_{\mathrm{B}}(\omega)+1|
        [\delta(\omega-\Omega)+\delta(\omega+\Omega)] \\
    R^{(\mathrm{Fr})}_{l}(\omega) & = & R'_{\mathrm{Fr}}
     \eta_{l} |n_{\mathrm{B}}(\omega)+1|
        [\delta(\omega-\Omega)+\delta(\omega+\Omega)]
\end{eqnarray*}
where
\begin{displaymath}
    R_{\mathrm{Fr}}\approx\frac{3\Omega e^{2}}%
        {8\pi\sqrt{2\pi}\epsilon_{0}\tilde{\epsilon}L}
        \frac{(l_{\mathrm{e}}^{2}-l_{\mathrm{h}}^{2})^{2}}{16L^{4}},\;\;\;
    R'_{\mathrm{Fr}}=\frac{\Omega e^{2}}%
        {8\pi\sqrt{2\pi}\epsilon_{0}\tilde{\epsilon}l_{\mathrm{h}}},
\end{displaymath}
and $\eta_{(00)}=1$, $\eta_{(01)}=1/4$, $\eta_{(02)}=3/32$,
$\eta_{10}=3/16$. Here we numbered the states by $(N,M)$, where
$M$ is the total angular momentum and $N$ is another quantum
number, and introduced the averaged carrier localization
$L^{2}=(l_{\mathrm{e}}^{2}+l_{\mathrm{h}}^{2})/2$, where
$l_{\mathrm{e}},l_{\mathrm{h}}$ correspond to $l_{\bot}$ in Eq.
(\ref{ground-state}) for electrons and holes, respectively, and
$l_{\mathrm{e}}^{2}-l_{\mathrm{h}}^{2}\ll L^{2}$.

\section{Decoherence of carrier states by a dressing process}
\label{sec:dress}

Any optical experiment or optical control of carrier states in a
quantum dot is based on the coupling between the carriers and the
electric field of the electromagnetic wave. By using high-power
laser pulses this coupling can be made strong, leading to fast
excitation of confined excitons. However, this coupling involves
pure electronic degrees of freedom, inducing only carrier
transitions which must then be followed by lattice relaxation to
the potential minimum corresponding to the newly created charge
state. In the case of the optical exciton creation, the initial
lattice configuration corresponds to zero polarization field and
zero deformation, while the eigenstate of the interacting system
involves some lattice polarization (LO phonon field) and some
deformation (LA phonon field). Thus, the process of the lattice
relaxation may be viewed as the formation of the phonon dressing
around the confined exciton. The lattice relaxation process,
accompanied by a non-zero probability of phonon emission, only
takes place if the result of a control pulse is a one-exciton
state. Hence, if the pulse creates a superposition, the lattice
dynamics is different for the two components of the superposition
state, leading to entanglement between the carrier and lattice
degrees of freedom and to decoherence. Physically, detection of a
phonon carrying out the excess energy during the lattice
relaxation process reduces the superposition of carrier states to
the one-exciton state, since the other state (empty dot) is never
accompanied by the lattice relaxation. It should be kept in mind,
however, that unlike in the phonon-assisted carrier relaxation
processes, the average number of phonons emitted during the
dressing is much less then one (in the weak coupling case) and
even in the limit of $t\to\infty$ the emission probability is
below 1, leading only to partial loss of the carrier coherence
\cite{krummheuer02,jacak03b}.

In this section we discuss the carrier-phonon kinetics after an
ultra-fast excitation from the point of view of decoherence and
``information leakage'' from the carrier subsystem \cite{jacak03b}
(see \cite{krummheuer02,vagov02a,vagov03} for an exhaustive
discussion from the point of view of the optical properties).

We consider a quantum dot in which an exciton is created by a very
short laser pulse. We assume that the pulse is much shorter than
the periods of the phonons but long enough to ensure spectral
overlap only with one (ground-state) exciton transition. We allow,
however, for the existence of dark states of energy much lower
than the closest bright state, as discussed in the section
\ref{sec:model}. In accordance with the discussion in section
\ref{sec:coupling}, we include in the interaction Hamiltonian
(\ref{H-int}) the deformation potential coupling to LA phonons and
Fr\"ohlich coupling to LO phonons.

In order to quantify the quality of a manipulation on the quantum
state from the point of view of the desired goal, one defines the
fidelity as the overlap between the desired final state obtained
from the initial state $|\psi_{0}\rangle$ by the unperturbed
evolution $U_{\mathrm{C}}$ and the actual state described by the
reduced density matrix of the carrier subsystem $\rho(t)$,
\begin{equation}
F=1-\delta=\langle \psi_{0} |U_{\mathrm{C}}^{\dagger}(t)
\rho(t)U_{\mathrm{C}}(t)|\psi_{0}\rangle_{T}, \label{fidelity}
\end{equation}
where $\langle\cdot\rangle_{T}$ denotes thermal average over the
initial lattice states. If the initial carrier state is the vacuum
$|0\rangle$ and the operation, performed at $t=0$, is an
ultra-fast rotation by the angle $\alpha$ on the Bloch sphere
corresponding to the $|0\rangle$ and $|1\rangle$ states than the
unperturbed state for $t>0$ is
\begin{equation}
    |\psi(t)\rangle=U_{\mathrm{C}}(t)|0\rangle
        =\cos\frac{\alpha}{2}|0\rangle
        -ie^{-iE_{1}t}\sin\frac{\alpha}{2}|1\rangle.
\label{free-evol}
\end{equation}
Let us calculate the actual state, including phonon perturbations,
in this ultrafast limit. The diagonal elements of the density
matrix remain constant, since phonon processes cannot change the
excitonic occupations,
\begin{displaymath}
    \rho_{00}=\cos^{2}\frac{\alpha}{2},\;\;\;
    \rho_{11}=\sin^{2}\frac{\alpha}{2},
\end{displaymath}
while for the non-diagonal ones one finds
\begin{equation}
    \rho_{10}(t) =
        {\mathrm{Tr}}\left[ \varrho(t)|0\rangle\!\langle 1|\right]
        =-\frac{1}{2}i\sin\alpha G(t), \label{densmat}
\end{equation}
where we have substituted
$\rho(0^{+})=|\psi(0)\rangle\!\langle\psi(0)| \otimes
w_{\mathrm{L}}$, with $|\psi(0)\rangle$ given by
(\ref{free-evol}), and defined the correlation function
\begin{displaymath}
    G(t)=\left\langle U^{\dag}(t)|0\rangle\!\langle 1|U(t)
        |1\rangle\!\langle 0| \right\rangle_{T},
\end{displaymath}
where $U(t)$ is the evolution operator for the total interacting
system. Using the definition (\ref{fidelity}), the fidelity loss
$\delta$ may be expressed as
\begin{equation}\label{delta-Green}
\delta(t)=\frac{1}{2}\sin^{2}\alpha\left[ 1-\mathrm{Re} \left(
G(t)e^{iE_{1}t} \right) \right].
\end{equation}

In order to find the correlation function we describe the
carrier-phonon kinetics after an abrupt excitation using the basis
of perturbative eigenstates of the Hamiltonian
$H=H_{\mathrm{C}}+H_{\mathrm{int}}$ \cite{jacak02a,jacak03a}.
Thus, we introduce the new states and operators
\begin{displaymath}
    |\tilde{n}\rangle=e^{\sf S}|n\rangle,\;\;\;
    \beta_{\bm{k}}=e^{\sf S}b_{\bf{k}}e^{-\sf S},
\end{displaymath}
corresponding to dressed particles. The system Hamiltonian
expressed in terms of these operators is diagonal up to
second-order corrections if the anti-hermitian operator $\sf S$ is
chosen as
\begin{displaymath}
    {\sf S}=\sum_{n,n'\ge 1;\bm{k},s}\Phi^{(s)}_{nn'}(\bm{k})
    |n\rangle\!\langle n'|
    (b_{\bm{k},s}-b_{-\bm{k},s}^{\dagger}),
\end{displaymath}
where
\begin{displaymath}
    \Phi^{(s)}_{nn'}(\bm{k})=
        \frac{F_{nn'}^{(s)}(\bm{k})}%
            {E_{n'}-E_{n}+\hbar\omega_{s}(\bm{k})},
\end{displaymath}
and $F_{nn'}^{(s)}$ are given by (\ref{cpl-X-1}--\ref{cpl-X-4}).
Therefore, the evolution of these new states and operators is
free,
\begin{displaymath}
\alpha_{n}(t) = |\tilde{n}\rangle= e^{-i
\frac{\tilde{E}_{n}}{\hbar}t}|\tilde{n}\rangle,\;\;\;
\beta_{s,\bm{k}}(t) = \beta_{s,\bm{k}}e^{-i \omega_{s}(\bm{k})t},
\end{displaymath}
where $\tilde{E}_{n}$ is the exciton energy including corrections
resulting from the interaction.

To the lowest order, the creation operator for the ground exciton
state may now be written as
\begin{eqnarray*}
    |1\rangle\!\langle 0| & = &
        e^{\sf S}|\tilde{1}\rangle\!\langle0| e^{-\sf S} \\
    & = & \left(
        1-\frac{1}{2N}\sum_{n,s,\bm{k}}|\phi_{0n}^{(s)}(\bm{k})|^{2}
        (2\beta_{s,\bm{k}}^{\dagger}\beta_{s,\bm{k}}+1) \right)
        |\tilde{1}\rangle\!\langle0| \\
    & & +\frac{1}{\sqrt{N}}\sum_{n,s,\bm{k}}\phi_{0n}^{(s)*}(\bm{k})
        |\tilde{n}\rangle\!\langle0| \left( \beta_{s,\bm{k}}^{\dagger}+
        \beta_{s,-\bm{k}} \right).
\end{eqnarray*}

Using these formulae, it is easy to write down the correlation
function
\begin{eqnarray}\label{Gt}
G(t) & = & \left(1- \frac{1}{N}
  \sum_{n,s,\bm{k}}|\phi_{0n}^{(s)}(\bm{k})|^{2}(2n_{s,\bm{k}}+1)
\right) e^{-i \frac{\tilde{E}_{1}}{\hbar}t} \nonumber\\
& &
+\frac{1}{N}\sum_{n,s,\bm{k}}n_{s,\bm{k}}|\phi_{0n}^{(s)}(\bm{k})|^{2}
e^{-i(\tilde{E}_{n}/\hbar-\omega_{s}(\bm{k}))t} \nonumber\\
& &
+\frac{1}{N}\sum_{n,s,\bm{k}}(n_{s,\bm{k}}+1)|\phi_{n0}^{(s)}(\bm{k})|^{2}
e^{-i(\tilde{E}_{n}/\hbar+\omega_{s}(\bm{k}))t}, \nonumber
\end{eqnarray}
where $n_{s,\bm{k}}$ are the thermal occupation numbers for phonon
modes. It should be noted that the perturbative treatment is valid
only when the above correlation function remains close to one. For
a fixed strength of interaction this requires a low enough
temperature (in practice, for a typical InAs/GaAs dot, this means
$T<100$ K).

In order to explain the time evolution of this function, let us
remember that the functions $\Phi_{nn'}^{(s)}(\bm{k})$ contain the
formfactors (\ref{formfactor}), effectively selecting a certain
wavenumber range $k_{0}^{(s)}\pm \frac{1}{2}\Delta k^{(s)}$,
centered around $k_{0}^{(s)}$ for each branch of phonons ($\Delta
k^{(s)}\sim 1/l$, where $l$ is the dot size). This corresponds to
a certain frequency range $\omega_{0}^{(s)}\pm\frac{1}{2}\Delta
\omega^{(s)}$, around a central frequency $\omega_{0}^{(s)}$. The
phases that enter in the summation in (\ref{Gt}) at the time $t$
spread effectively over the angle range $\omega_{0}^{(s)}t\pm
\frac{1}{2}\Delta\omega^{(s)}t$. When this phase spreading reaches
$2\pi$, i.e. for $t\sim \frac{2\pi}{\Delta \omega^{(s)}}$, the
last two terms in (\ref{Gt}) become   small. The asymptotic value
of the corresponding error is
\begin{displaymath}
\delta= \frac{1}{2}\sin^{2}\alpha
\frac{1}{N}\sum_{\bm{k},n,s}|\phi_{0n}^{(s)}(\bm{k})|^{2}
(2n_{s,\bm{k}}+1).
\end{displaymath}

\begin{figure}[tb]
\unitlength 1mm
\begin{center}
\begin{picture}(135,28)(0,13)
\put(0,0){\resizebox{135mm}{!}{\includegraphics{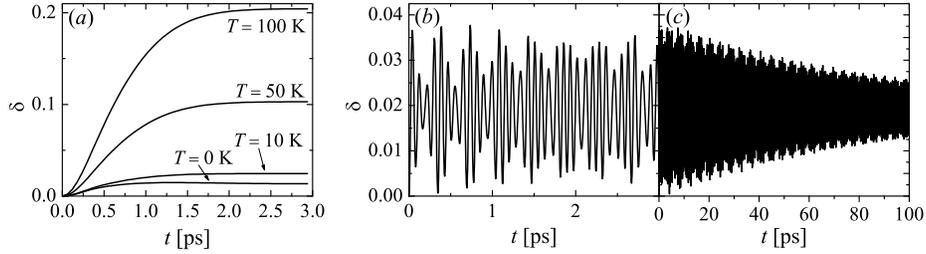}}}
\end{picture}
\end{center}
\caption{\label{fig:dress}Dressing-induced decoherence of the
exciton due to LA phonons (a) and due to LO phonons: oscillations
with LO phonon frequency (b) decay on a very long timescale (c)
(anharmonicity effects are not included). Four lowest exciton
states calculated by numerical diagonalization
\protect\cite{jacak03b} were taken into account.}
\end{figure}

In the case of acoustical phonons this value critically depends on
temperature (Fig. \ref{fig:dress}a). The decoherence time depends
on the phonon dispersion (the frequency range
$\Delta\omega^{(s)}$) and weakly evolves with temperature. For the
acoustical phonons, $\tau\sim 2\pi l/c\sim 1$ ps (Fig.
\ref{fig:dress}a). For the nearly dispersionless optical
pho\-nons, the dynamics is dominated by slowly vanishing coherent
phonon beats (Fig. \ref{fig:dress}b,c). However, in order to
excite these oscillations one needs pulses shorter than LO phonon
periods, i.e. of durations $\sim 10$ fs. Moreover, it is known
that the anharmonic LO--TA interaction \cite{vallee91,vallee94}
(not included in the present description) acts on much shorter
timescales and may be expected to considerably shorten this time.

\begin{figure}[tb]
\unitlength 1mm
\begin{center}
\begin{picture}(60,40)(0,5)
\put(0,0){\resizebox{60mm}{!}{\includegraphics{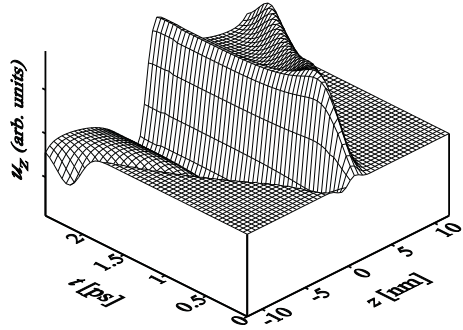}}}
\end{picture}
\end{center}
\caption{\label{fig:deform}Time evolution of the mean lattice
deformation (acoustic phonons) at the dot axis ($x=y=0$). The
time-independent deformation, corresponding to the coherent phonon
dressing, is formed around $z=0$ (i.e. in the dot area) within
$\approx 1$ ps. This is accompanied by emission of a phonon packet
carrying away the excess energy at the speed of sound, seen on the
plot as the ridge and the valley with growing distance from the
dot.}
\end{figure}

It is possible to see directly that the described process consists
in the formation of a coherent (i.e. with non-vanishing mean
displacement) phonon field corresponding to the classical lattice
deformation around the confined charge. Let us consider the mean
lattice displacement corresponding to the branch $s$ at the point
$\bm{r}$ after time $t$ after rapid creation of an exciton. Using
(\ref{phonon}) and transforming to the dressed basis one obtains
in the lowest order
\begin{displaymath}
\left\langle 1 |\bm{u}_{s}(\bm{r},t) |0 \right\rangle_{T} =
\frac{1}{N}\sum_{\bm{k}}\sqrt{\frac{\hbar}{2\varrho v\omega_{s}(\bm{k})}} \\
\frac{\bm{k}}{k}e^{i\bm{k}\cdot\bm{r}} 2\mathrm{Re}\left[
\phi_{00}^{s*}(\bm{k})\left( 1-e^{i\omega_{s}(\bm{k})t} \right)
\right].
\end{displaymath}
After a sufficiently long time the oscillating term averages to
zero (around $\bm{r}=0$) and a time-independent displacement field
is formed. The Fig. \ref{fig:deform} shows the mean lattice
deformation due to the deformation potential coupling on the dot
axis, $\bm{r}=(0,0,z)$, at various instances of time. A simple and
intuitive picture emerges: formation of the lattice deformation,
corresponding to the displaced equilibrium, is accompanied by
emitting a phonon packet that carries the excess energy away from
the dot with the speed of sound. It is clear that as soon as these
wavepackets leave the volume of the dot, some information about
the carrier state is carried into the outside world. Also, since
the only controlled interaction is that with the confined
carriers, once the phonons have been radiated out the
reversibility is lost.

We have shown that the carrier-phonon coupling leads to
entanglement of the carrier system and the surrounding lattice.
This, in turn, reduces the degree of coherence of the carrier
subsystem. This decoherence is related to the spontaneous
relaxation of the lattice to the new equilibrium defined by the
carrier-phonon interaction and results from the non-adiabaticity
of carrier evolution with respect to the lattice response times.
It may be expected that for a slow enough evolution most of the
lattice modes will follow adiabatically, thus creating the
coherent dressing cloud in a reversible way. In the following
sections we derive the formalism suitable for the description of
the carrier-phonon kinetics and apply it to a few problems.

\section{Evolution of the density matrix for a
driven carrier-phonon system}
\label{sec:method}

In this section we derive the equations for the reduced density
matrix of the carrier subsystem in the leading order in the phonon
coupling, assuming that the unperturbed evolution is known. This
somewhat lengthy and technical derivation yields very simple and
intuitive formulas that may be easily applied to a range of
problems to be discussed in the subsequent sections.

We will consider a system composed of charge carriers (electrons
and holes), localized in one or more quantum dots, driven
externally and interacting with phonons, as described by the
Hamiltonian (\ref{ham-gener}). As already mentioned, the evolution
of the carrier subsystem is generated by the (time dependent)
Hamiltonian $H_{\mathrm{C}}$, describing the properties of the
system itself as well as its coupling to driving fields. In this
chapter, we deal only with optical driving (although not
necessarily via direct excitation of dipole-allowed transitions),
but other types of external fields may be treated on the same
footing. The evolution of the phonon subsystem (reservoir) is
described by the Hamiltonian $H_{\mathrm{ph}}$ [Eq. (\ref{H-ph})].
We will restrict the discussion to the free phonon evolution,
neglecting anharmonicity effects. The evolution operator for the
driven carrier subsystem and free phonon modes, without
carrier-phonon interaction is
\begin{displaymath}
    U_{0}(t)=U_{\mathrm{C}}(t)\otimes e^{-iH_{\mathrm{ph}}(t-s)},
\end{displaymath}
where $U_{\mathrm{C}}(t)$ is the operator for unperturbed
evolution of the carrier subsystem (we suppress the initial time
in $U(t,s)$).

The carrier-phonon coupling is described by the interaction
Hamiltonian (\ref{H-int}) which may be written in the form
\begin{equation}\label{int}
    V=\sum_{nn'}S_{nn'}\otimes R_{nn'},
\end{equation}
where $n,n'$ run over the carrier subsystem states, $S_{nn'}$
(possibly time-dependent) act in the Hilbert space of the carrier
subsystem while the time-independent $R_{nn'}$ affect only the
environment. It is convenient to allow non-Her\-miti\-an operators
$S_{nn'}$ and $R_{nn'}$; however, we demand the symmetry relation
\begin{equation}\label{sym}
    S^{\dag}_{nn'}=S_{n'n},\;\;\;R_{nn'}^{\dag}=R_{n'n}
\end{equation}
which guarantees hermicity of the Hamiltonian (\ref{int}) and is
also explicitly satisfied by (\ref{H-int}).

We will assume that at the initial time $s$ the system is in the
product state
\begin{equation}\label{init}
    \varrho(s)=|\psi_{0}\rangle\!\langle\psi_{0}|\otimes\rho_{T},
\end{equation}
where $|\psi_{0}\rangle$ is a certain state of the carrier
subsystem and $\rho_{T}$ is the thermal equilibrium distribution
of phonon modes. Physically, such an assumption is usually
reasonable due to the existence of two distinct time scales: the
long one for the carrier decoherence (e.g. 1 ns ground state
exciton lifetime \cite{borri01,bayer02}) and the short one for the
reservoir relaxation (1 ps dressing time
\cite{borri01,krummheuer02,jacak03b}).

The starting point is the evolution equation for the density
matrix of the total system in the interaction picture with respect
to the externally driven evolution $U_{0}$, in the second order
Born approximation with respect to the carrier-phonon interaction
\cite{cohen98}
\begin{equation}\label{evol0}
    \tilde{\varrho}(t)=\tilde{\varrho}(s)
    +\frac{1}{i\hbar}\int_{s}^{t}d\tau[V(\tau),\varrho(s)]
    -\frac{1}{\hbar^{2}}\int_{s}^{t}d\tau\int_{s}^{\tau}d\tau'
      [V(\tau'),[V(\tau''),\varrho(s)]],
\end{equation}
where
\begin{displaymath}
    \tilde{\varrho}(t)=U_{0}^{\dag}(t)\varrho(t)U_{0}(t),\;\;\;
    V(t)=U_{0}^{\dag}(t)VU_{0}(t)
\end{displaymath}
(it should be kept in mind that $V$ may depend on time itself).

The reduced density matrix of the carrier subsystem at time $t$ is
\begin{displaymath}
    \rho(t)=U_{\mathrm{C}}(t)\tilde{\rho}(t)U_{\mathrm{C}}^{\dag}(t),\;\;\;
    \tilde{\rho}(t)=\mathrm{Tr}_{\mathrm{R}}\tilde{\varrho}(t),
\end{displaymath}
where the trace is taken over the reservoir degrees of freedom.
The first (zeroth order) term in (\ref{evol0}) obviously gives
rise to
\begin{equation}\label{ro0}
    \rho^{(0)}(t)
        =U_{\mathrm{C}}(t)|\psi_{0}\rangle\!\langle\psi_{0}|
            U_{\mathrm{C}}^{\dag}(t)
        =|\psi_{0}(t)\rangle\!\langle\psi_{0}(t)|.
\end{equation}
The second term vanishes, since it contains the thermal average of
an odd number of phonons. The third (second order) term describes
the leading phonon correction to the dynamics of the carrier
subsystem,
\begin{equation}\label{ro2}
    \tilde\rho^{(2)}(t)=
    -\frac{1}{\hbar^{2}}\int_{s}^{t}d\tau\int_{s}^{\tau}d\tau'
      \mathrm{Tr}_{\mathrm{R}}[V(\tau'),[V(\tau''),\varrho(s)]].
\end{equation}

First of the four terms resulting from expanding the commutators
in (\ref{ro2}) is
\begin{displaymath}
    (\mathrm{I})=-Q_{t}|\psi_{0}\rangle\!\langle\psi_{0}|,
\end{displaymath}
where
\begin{equation}\label{Q}
    Q_{t}=\frac{1}{\hbar^{2}}\sum_{nn'}\sum_{mm'}
    \int_{s}^{t}d\tau\int_{s}^{\tau}d\tau'
    S_{nn'}(\tau)S_{mm'}(\tau')
     \langle R_{nn'}(\tau-\tau')R_{mm'}\rangle.
\end{equation}
The operators $S$ and $R$ are transformed into the interaction
picture in the usual way
\begin{displaymath}
    S_{nn'}(t)=U_{0}^{\dag}(t)S_{nn'}U_{0}(t),\;\;\;
    R_{nn'}(t)=U_{0}^{\dag}(t)R_{nn'}U_{0}(t)
\end{displaymath}
and $\langle\hat\mathcal{O}\rangle
=\mathrm{Tr}_{\mathrm{R}}[\hat\mathcal{O}\rho_{T}]$ denotes the
thermal average (obviously $[U_{0}(t),\rho_{T}]=0$).

The second term is
\begin{displaymath}
    (\mathrm{II})= \frac{1}{\hbar^{2}}\sum_{nn'}\sum_{mm'}
    \int_{s}^{t}d\tau\int_{s}^{\tau}d\tau'
    |\psi_{0}\rangle\!\langle\psi_{0}|
    S_{mm'}(\tau')S_{nn'}(\tau)
    \langle R_{mm'}(\tau'-\tau)R_{nn'}\rangle.
\end{displaymath}
Using the symmetry relations (\ref{sym}) one has
\begin{equation}\label{sym2}
    \langle R_{mm'}(\tau'-\tau)R_{nn'}\rangle^{*}
    =\langle R_{n'n}(\tau-\tau')R_{m'm}\rangle,\;\;\;
    \left[S_{mm'}(\tau')S_{nn'}(\tau) \right]^{\dag}
    =S_{n'n}(\tau)S_{m'm}(\tau'),
\end{equation}
hence this term may be written as
\begin{displaymath}
    (\mathrm{II})=-|\psi_{0}\rangle\!\langle\psi_{0}|Q^{\dag}_{t}.
\end{displaymath}

In a similar manner, using the symmetries (\ref{sym}), the two
other terms may be combined to
\begin{displaymath}
    (\mathrm{III})+(\mathrm{IV})=
    \hat{\Phi}_{t}\left[|\psi_{0}\rangle\!\langle\psi_{0}|\right].
\end{displaymath}
where
\begin{equation}\label{Phi}
    \hat{\Phi}_{t}\left[\rho\right]=
    \frac{1}{\hbar^{2}}\sum_{nn'}\sum_{mm'}
    \int_{s}^{t}d\tau\int_{s}^{\tau}d\tau'
    S_{nn'}(\tau')\rho S_{mm'}(\tau)
     \langle R_{mm'}(\tau-\tau')R_{nn'}\rangle.
\end{equation}

In terms of the new Hermitian operators
\begin{equation}\label{Ah}
    A_{t}=Q_{t}+Q_{t}^{\dag},\;\;\;
    h_{t}=\frac{1}{2i}(Q_{t}-Q^{\dag}_{t}),
\end{equation}
the density matrix at the final time $t$ (\ref{ro0},\ref{ro2}) may
be written as
\begin{equation}\label{master}
    \rho(t)=U_{\mathrm{C}}(t)\left(|\psi_{0}\rangle\!\langle\psi_{0}|
    -i\left[ h_{t},|\psi_{0}\rangle\!\langle\psi_{0}| \right]
    -\frac{1}{2}\left\{A_{t},|\psi_{0}\rangle\!\langle\psi_{0}|\right\}
    +\hat{\Phi}_{t}[|\psi_{0}\rangle\!\langle\psi_{0}|]\right)
    U^{\dag}_{\mathrm{C}}.
\end{equation}
The first term is a hamiltonian correction which does not lead to
irreversible effects and, in principle, may be compensated for by
an appropriate modification of the control Hamiltonian
$H_{\mathrm{C}}$. The other two terms describe processes of
entangling the system with the reservoir, leading to the loss of
coherence of the carrier state.

Let us introduce the spectral density of the reservoir,
\begin{equation}\label{spdens}
    R_{nn',mm'}(\omega)=\frac{1}{2\pi\hbar^{2}}\int dt
    \langle R_{nn'}(t)R_{mm'}\rangle e^{i\omega t}.
\end{equation}
If the operators $R_{nn'}$ are linear combinations of freely
evolving bosonic modes (only this case will be considered in this
chapter),
\begin{equation}\label{Rnn}
    R_{nn'}=R_{n'n}^{\dag} =\frac{1}{\sqrt{N}}\sum_{\bm{k}}
    F_{nn'}(\bm{k})\left(b_{\bm{k}}+b_{-\bm{k}}^{\dag}\right),
\end{equation}
with $F_{nn'}(\bm{k})=F_{n'n}^{*}(-\bm{k})$ (branch index implicit
in $\bm{k}$), then (\ref{spdens}) coincides with
(\ref{spdens-expli}).

With the help of (\ref{spdens}) one may write
\begin{equation}\label{Fi}
    \hat{\Phi}_{t}\left[\rho\right]=
    \sum_{nn'}\sum_{mm'}
    \int d\omega R_{nn',mm'}(\omega)
    Y_{mm'}(\omega)\rho Y_{n'n}^{\dag}(\omega)
\end{equation}
where the frequency-dependent operators have been introduced,
\begin{equation}\label{Y}
    Y_{nn'}(\omega)
        =\int_{s}^{t}d\tau S_{nn'}(\tau)e^{i\omega \tau}.
\end{equation}
Using (\ref{spdens}) one has also
\begin{displaymath}
    Q_{t}=
    \sum_{nn'}\sum_{mm'}d\omega\int_{s}^{t}d\tau\int_{s}^{t}d\tau'
    \theta(\tau-\tau')S_{nn'}(\tau)S_{mm'}(\tau')
    R_{nn',mm'}(\omega)e^{-i\omega(\tau-\tau')}.
\end{displaymath}
Next, representing the Heaviside function as
\begin{displaymath}
    \theta(t)=-e^{i\omega t}\int\frac{d\omega'}{2\pi i}
    \frac{e^{-i\omega' t}}{\omega'-\omega+i0^{+}},
\end{displaymath}
we write
\begin{eqnarray*}
    Q_{t} & = & -\sum_{nn'}\sum_{mm'}\int d\omega R_{nn',mm'}(\omega)
    \int\frac{d\omega'}{2\pi i}
    \frac{Y_{n'n}^{\dag}(\omega')Y_{mm'}(\omega')}{\omega'-\omega+i0^{+}} \\
    & = & - \sum_{nn'}\sum_{mm'}\int d\omega R_{nn',mm'}(\omega) \\
  & &   \times
    \int\frac{d\omega'}{2\pi i} Y_{n'n}^{\dag}(\omega')Y_{mm'}(\omega')
    \left[-i\pi\delta(\omega'-\omega)+\mathcal{P}\frac{1}{\omega'-\omega}
    \right],
\end{eqnarray*}
where $\mathcal{P}$ denotes the principal value.

Hence, the two Hermitian operators defined in (\ref{Ah}) take the
form
\begin{equation}\label{A}
    A_{t}=
    \sum_{nn'}\sum_{mm'}\int d\omega
    R_{nn',mm'}(\omega)Y_{n'n}^{\dag}(\omega)Y_{mm'}(\omega)
\end{equation}
and
\begin{equation}\label{h}
    h_{t}=
    \sum_{nn'}\sum_{mm'}
    \int d\omega R_{nn',mm'}(\omega)
    \mathcal{P} \int\frac{d\omega'}{2\pi}
    \frac{Y_{n'n}^{\dag}(\omega')Y_{mm'}(\omega')}{\omega'-\omega},
\end{equation}
where we have used the relation
\begin{displaymath}
    R^{*}_{nn',mm'}(\omega)=R_{m'm,n'n}(\omega),
\end{displaymath}
resulting from the definition (\ref{spdens}) and the symmetry
relation (\ref{sym2}).

Using the definition of the fidelity (\ref{fidelity}) and the
Master equation (\ref{master}), the error may be written in a
general case as
\begin{displaymath}
    \delta=\left\langle \psi_{0}\left|A_{t}\right|\psi_{0}\right\rangle
      -\left\langle \psi_{0}\left|
            \hat{\Phi}\left[|\psi_{0}\rangle\!\langle\psi_{0}|
                \right]\right|
                \psi_{0}\right\rangle.
\end{displaymath}
It should be noted that the unitary correction generated by
$h_{t}$ does not contribute to the error at this order.

Using the definitions (\ref{Phi},\ref{A}) this may be further
transformed to
\begin{equation}\label{delta-detal}
    \delta=\sum_{nn'mm'}\int d\omega R_{nn',mm'}(\omega)
    \left\langle \psi_{0}\left|Y_{n'n}^{\dag}(\omega)\mathsf{P}^{\bot}
    Y_{mm'}(\omega)\right|\psi_{0} \right\rangle,
\end{equation}
where $\mathsf{P}^{\bot}$ is the projector on the orthogonal
complement of $|\psi_{0}\rangle$ in the carrier space.

\section{Phonon-induced damping of Rabi oscillations}
\label{sec:rabi}

\paragraph{Introduction.}
Observation of Rabi oscillations of charge degrees of freedom
confined in quantum dots
\cite{stievater01,kamada01,zrenner02,borri02a,htoon02} is believed
to be a fundamental step towards quantum control of these systems.
However, the coherent dynamics of the confined carrier states is
very sensitive to any interaction with the macroscopic number of
degrees of freedom of the outside world. In fact, so far it has
always turned out that experimentally observable Rabi oscillations
deviate from the ideal ones, the discrepancy being larger for
stronger pulses. In principle, this might also be explained by
experimental conditions or environmental perturbation: scattering
by weakly localized excitons around an interface fluctuation QD
(further confirmed by increasing decay for stronger pulses)
\cite{stievater01}, tunneling to leads in the photodiode structure
(on $\sim 10$ ps timescale) \cite{zrenner02}, or dipole moment
distribution in the QD ensemble \cite{borri01}.

One might believe that all the perturbation comes from sources
that may be removed or minimized by technology improvement and by
optimizing the experimental conditions and hence produce no
fundamental obstacle to arbitrarily perfect quantum control over
the excitonic states. However, in every case the QDs are
inherently coupled to the surrounding crystal lattice. The recent
theoretical study \cite{forstner03} on optical Rabi flopping of
excitons in QDs driven by finite-length optical pulses in the
presence of the lattice reservoir shows that exponential damping
models fail to correctly describe the system kinetics. The
appropriate quantum-kinetic description yields much less damping,
especially for long pulses. It turns out that the lowest
``quality'' of the Rabi oscillation is obtained for pulse
durations of a few ps, while for longer durations the damping is
again decreased.

In this section we discuss both qualitative and quantitative
explanation \cite{machnikowski03a} of the mechanism leading to the
phonon-induced damping reported in the theoretical and
experimental studies. We show that the carrier-phonon interaction
responsible for the damping of the oscillations has a resonant
character: While in the linear limit the system response depends
only on the spectral decomposition of the pulse, the situation is
different when a strong pulse induces an oscillating charge
distribution in the system. In a semi-classical picture, this
would act as a driving force for the lattice dynamics. If the
induced carrier dynamics is much faster than phonon oscillations
the lattice has no time to react until the optical excitation is
done. The subsequent dynamics will lead to exciton dressing,
accompanied by emission of phonon packets, and will partly destroy
coherence of superposition states
\cite{borri01,krummheuer02,jacak03b,vagov02a} but cannot change
the exciton occupation number. In the opposite limit, the carrier
dynamics is slow enough for the lattice to follow adiabatically.
The optical excitation may then be stopped at any stage without
any lattice relaxation incurred, hence with no coherence loss. The
intermediate case corresponds to modifying the charge distribution
in the QD with frequencies resonant with the lattice modes which
leads to increased interaction with phonons and to decrease of the
carrier coherence (see Ref. \cite{axt99} for a simple, single-mode
model).

\paragraph{The formalism}
In one version of the experiment
\cite{stievater01,zrenner02,borri02a} one measures the average
occupation of the exciton ground state, $\langle
1|\rho(\infty)|1\rangle$, after a resonantly coupled pulse of
fixed length but variable amplitude, starting with the system in
the ground state $|0\rangle$. According to (\ref{free-evol}), in
the ideal case the final occupation should vary as
\begin{displaymath}
    \langle 1|\rho(\infty)|1\rangle_{\mathrm{ideal}}
        =\sin^{2}\frac{1}{2}\alpha,
\end{displaymath}
where
\begin{displaymath}
    \alpha=\Phi(\infty)=\int_{-\infty}^{\infty}d\tau f(\tau),
\end{displaymath}
is the total pulse area. Within our formalism, the final
occupation of the exciton state is given directly by the
appropriate diagonal element of (\ref{master}), once the spectral
integrals entering in (\ref{Phi}) and (\ref{A}) for a given pulse
are calculated.

The interaction Hamiltonian has the form (\ref{H-int}), with the
coupling constants (\ref{cpl-X-1}--\ref{cpl-X-4}). Hence,
$S_{nn'}=|n\rangle\!\langle n'|$, $n=1,2,\ldots$ and, using the
explicit form of the evolution operator (\ref{free-evol}), the
operators (\ref{Y}) may be written in the form ($n,n'>1$; we
integrate by parts in order to extract the oscillatory
contributions)
\begin{eqnarray}
\label{Y11}
  Y_{11} = &=& (|0\rangle\!\langle1|-|1\rangle\!\langle0|)
    \frac{1}{2\omega}
        \left[ \sin\alpha e^{i\omega t}
            -K_{\mathrm{s}}^{(1)}(\omega) \right] \\
\nonumber
  & & -(|1\rangle\!\langle1|-|0\rangle\!\langle0|)\frac{i}{2\omega}
        \left[ \cos\alpha e^{i\omega t}-e^{i\omega s}
            -K_{\mathrm{c}}^{(1)}(\omega) \right] \\
\nonumber
  & & -(|1\rangle\!\langle1|+|0\rangle\!\langle0|)
        \frac{1}{2\omega}\left[e^{i\omega t}-e^{i\omega s}
            \right], \\
\label{Yn1}
  Y_{n1} &=& -|n\rangle\!\langle 0|\frac{1}{\omega+\epsilon_{n}}
            e^{-i\epsilon_{n}s}  \left[
            \sin\frac{\alpha}{2} e^{i(\omega+\epsilon_{n})t}
            -K_{\mathrm{s}}^{(1/2)}(\omega+\epsilon_{n}) \right] \\
\nonumber
  & &   -|n\rangle\!\langle 1|\frac{i}{\omega+\epsilon_{n}}
            e^{-i\epsilon_{n}s} \left[
            \cos\frac{\alpha}{2} e^{i(\omega+\epsilon_{n})t}
            -e^{i(\omega+\epsilon_{n})s}
            -K_{\mathrm{c}}^{(1/2)}(\omega+\epsilon_{n}) \right] \\
  Y_{nn'} &=& -|n\rangle\!\langle n'|
    \frac{i}{\omega+\epsilon_{n}-\epsilon_{n'}}
    e^{-i(\epsilon_{n}-\epsilon_{n'})s}
    \left[ e^{i(\omega+\epsilon_{n}-\epsilon_{n'})t}
    -e^{i(\omega+\epsilon_{n}-\epsilon_{n'})s} \right],\label{Ynn}
\end{eqnarray}
where $\alpha=\Phi(\infty)$ is the total rotation angle and
\begin{equation}\label{ks-kc}
K^{(\mu)}_{\mathrm{s}}(\omega) = \int_{s}^{t}d\tau e^{i\omega\tau}
\frac{d}{d\tau}\sin \mu\Phi(\tau),\;\;\;
K^{(\mu)}_{\mathrm{c}}(\omega) = \int_{s}^{t}d\tau e^{i\omega\tau}
\frac{d}{d\tau}\cos \mu\Phi(\tau).
\end{equation}
In the limit of $s\to-\infty$, $t\to\infty$ (i.e. measurement
after sufficiently long time compared to the reservoir memory),
the functions (\ref{ks-kc}) actually depend only on
$\omega\tau_{\mathrm{p}}$, where $\tau_{\mathrm{p}}$ is the pulse
duration.

Since the unperturbed evolution operator (\ref{free-evol}) has a
block-diagonal structure and conserves the subspace spanned by
$|0\rangle$ and $|1\rangle$, the value of $\rho_{11}(t)$ is
determined by the matrix elements of $\tilde{\rho}(t)$ between
these two states, while $\rho_{ll}$ for $l>1$ has the same form as
in the interaction picture. Retaining only non-vanishing
contributions, these matrix elements are, from (\ref{master}),
with $|\psi_{0}\rangle=|0\rangle$,
\begin{eqnarray}
\label{ro00}
  \tilde{\rho}_{00} &=& 1-\sum_{l\ge 1}\int d\omega
        R_{l}(\omega)|\langle l|Y_{l1}|0\rangle|^{2},\\
\label{ro01}
  \tilde{\rho}_{01} &=& \int d\omega R_{1}(\omega) \\
\nonumber
 & & \times
     \left(\langle 0|Y_{11}|0\rangle\!\langle 1|Y_{11}|0\rangle^{*}
        -\frac{1}{2}\langle 0|Y_{11}|0\rangle^{*}
            \langle 0|Y_{11}|1\rangle
        -\frac{1}{2}\langle 1|Y_{11}|0\rangle^{*}
            \langle 1|Y_{11}|1\rangle\right)\\
\nonumber
 & &  -\frac{1}{2} \sum_{l\ge 2}\int d\omega
        R_{l}(\omega)\langle l|Y_{l1}|0\rangle^{*}
            \langle l|Y_{l1}|1\rangle \\
 \label{roll}
  \tilde{\rho}_{ll} &=& \int d\omega
        R_{l}(\omega)|\langle l|Y_{l1}|0\rangle|^{2},\;\;\;l\ge 1,
\end{eqnarray}
where the short-hand notation $R_{l}(\omega)\equiv
R_{1l,l1}(\omega)$ has been introduced [the definition is given by
Eq. (\ref{spdens-expli})]. Apart from the dissipative correction
described by the above formulas, there is an additional component,
generated by the hermitian operator $h_{t}$ [cf. Eq.
(\ref{master})], describing a unitary correction, e.g. light
coupling renormalization and energy shifts, which may be canceled
by an appropriate modification of $H_{\mathrm{C}}$ (these effects
may lead e.g. to intensity dependence of the observed Rabi
frequency\cite{forstner03}).

The phonon spectral densities $R_{l}(\omega)$ are obtained from
(\ref{spdens-expli}). We assume that no phonon-induced transitions
are possible, i.e. the spectral densities $R_{l}(\pm\epsilon_{l})$
vanish. For any physical carrier-phonon coupling at arbitrary
temperature one has also $R(\omega)\sim\omega^{n}$, $n\ge 2$ for
$\omega\to 0$. Thus, after substituting (\ref{Y11}--\ref{Ynn})
into (\ref{ro00}--\ref{roll}) and taking the limit $s\to -\infty$,
$t\to\infty$ the contribution from the oscillating terms vanishes
and the relevant elements of the density matrix, upon transforming
back to the Schr\"odinger picture using (\ref{free-evol}), are
\begin{eqnarray}
  \tilde{\rho}_{00} &=& \cos^{2}\frac{\alpha}{2}-\int d\omega
      \frac{R_{1}(\omega)}{\omega^{2}}
        S_{\mathrm{a}}(\omega)
      -\sum_{l>1} \int d\omega
      \frac{R_{l}(\omega-\epsilon_{l})}{\omega^{2}}
        S_{\mathrm{b}}(\omega) \label{ro00S} \\
  \tilde{\rho}_{11} &=& \sin^{2}\frac{\alpha}{2}\left[
    1-\sum_{l>1}\int d\omega
        \frac{R_{l}(\omega-\epsilon_{l})}{\omega^{2}}
  \right] \\
 & &   + \int d\omega
     \frac{R_{1}(\omega)}{\omega^{2}} S_{\mathrm{a}}(\omega)
  -\sum_{l>1} \int d\omega
        \frac{R_{l}(\omega-\epsilon_{l})}{\omega^{2}}
            S_{\mathrm{c}}(\omega)        \label{ro11S}\\
  \tilde{\rho}_{ll} &=& \sin^{2}\frac{\alpha}{2} \int d\omega
        \frac{R_{l}(\omega-\epsilon_{l})}{\omega^{2}}
        + \int d\omega
     \frac{R_{1}(\omega-\epsilon_{l})}{\omega^{2}}
        \left[ S_{\mathrm{b}}(\omega) +S_{\mathrm{c}}(\omega)\right],
        \label{rollS}
\end{eqnarray}
where
\begin{eqnarray}
  S_{\mathrm{a}}(\omega) &=& \frac{1}{4}\left\{
        \cos\alpha |K_{\mathrm{s}}^{(1)}(\omega)|^{2}
        -\sin\alpha \mathrm{Re}\left[K_{\mathrm{s}}^{(1)*}(\omega)
            K_{\mathrm{c}}^{(1)}(\omega)\right] \right\}, \\
  S_{\mathrm{b}}(\omega) &=&
        \cos^{2}\frac{\alpha}{2}|K_{\mathrm{s}}^{(1/2)}(\omega)|^{2}
        -\frac{1}{2}\sin\alpha \mathrm{Re}\left[
        K_{\mathrm{s}}^{(1/2)*}(\omega)
            K_{\mathrm{c}}^{(1/2)}(\omega)\right]\\
  S_{\mathrm{c}}(\omega) &=&
        \sin^{2}\frac{\alpha}{2}|K_{\mathrm{s}}^{(1/2)}(\omega)|^{2}
        +\frac{1}{2}\sin\alpha \mathrm{Re}\left[
        K_{\mathrm{s}}^{(1/2)*}(\omega)
            K_{\mathrm{c}}^{(1/2)}(\omega)\right].
\end{eqnarray}

\begin{figure}[tb]
\unitlength 1mm
\begin{center}
\begin{picture}(135,30)(0,13)
\put(0,0){\resizebox{135mm}{!}{\includegraphics{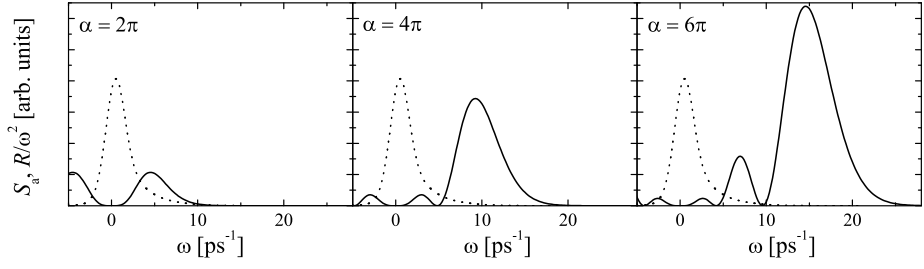}}}
\end{picture}
\end{center}
\caption{\label{fig:sr}Phonon spectral density
$R(\omega)/\omega^{2}$ (dotted) for deformation potential coupling
to LA phonons at $T=10$ K and the nonlinear pulse spectrum
$S_{\mathrm{a}}(\omega)$ (solid lines) for $\tau_{\mathrm{p}}=1$
ps, for rotation angles $\alpha$ as shown.}
\end{figure}

The formulas (\ref{ro00S}--\ref{ro11S}) quantify the idea of
resonance between the induced dynamics and lattice modes: in Fig.
\ref{fig:sr} the phonon spectral density is compared to the
non-linear frequency characteristics of the optically controlled
exciton dynamics for $\tau_{\mathrm{p}}=1$ ps (the characteristics
for other durations is easily obtained by scaling). According to
(\ref{ro00S}--\ref{ro11S}), the overlap of these spectral
characteristics with the phonon spectral density gives the
perturbation of the coherent carrier dynamics.

All the spectral densities $R_{l}(\omega-\epsilon_{l})$ contain
contributions from all the phonon branches and all carrier--phonon
coupling channels. For the LO phonons, they are peaked at
$\omega=\epsilon_{l}+\Omega$ (at low temperatures). One should
note that the denominators $\Omega$ and $\Omega+\epsilon_{l}$ are
of the same order of magnitude, while the strength of the spectral
density is much stronger for $l\ge 2$ because of the charge
cancelation effect decreasing the ground-state contribution.
Therefore, the response form the higher states is stronger.
Physically, it corresponds to phonon-assisted excitation of a dark
state. Obviously, the spectral functions
$S_{\mathrm{a-c}}(\omega)$ must extend to this high frequency
sector.

For acoustical phonons, in contrast, the spectral density are
concentrated at low frequencies, so that the denominators
$(\omega+\epsilon_{l})^{2}$ for $l\ge 2$ are at least two orders
of magnitude higher than the typical frequency $\omega^{2}$
appearing in the $l=1$ term. Therefore, the contribution from the
acoustic phonons is restricted to the ground state term.  For the
piezoelectric coupling, this term is strongly reduced by the
charge cancelation; therefore, the effect of this coupling on the
exciton coherence may be neglected as long as the electron and
hole densities for the excitonic ground state overlap.

\begin{figure}[tb]
\unitlength 1mm
\begin{center}
\begin{picture}(135,30)(0,13)
\put(0,0){\resizebox{135mm}{!}{\includegraphics{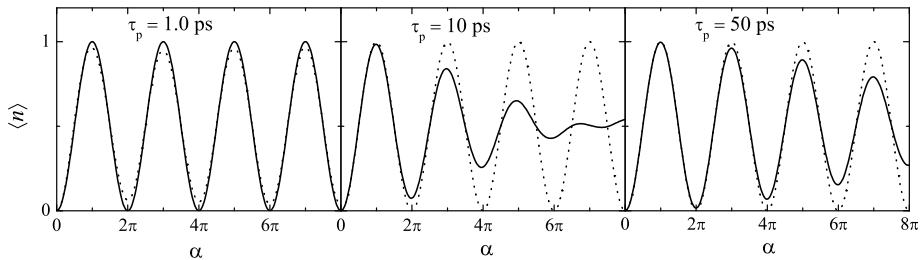}}}
\end{picture}
\end{center}
\caption{\label{fig:rabi}Pulse-area-dependent Rabi oscillations
for various pulse durations $\tau_{\mathrm{p}}$ as shown in the
figure, for $T=10$ K ($\alpha$ is the rotation angle on the Bloch
sphere). Dotted line shows unperturbed oscillations. The
wavefunction localization widths are $l_{\mathrm{e}}=4.9$ nm,
$l_{\mathrm{h}}=4.0$ nm, $l_{z}=1$ nm.}
\end{figure}

\paragraph{Damping due to the LA phonons.}
Let us first discuss the case of relatively long pulses
(picosecond durations) and discuss the contributions from the LA
phonons only, neglecting the existence of the higher states. The
results of such calculations is shown in Fig. \ref{fig:rabi} for
Gaussian pulses [$\tau_{\mathrm{p}}$ is the full width at half
maximum of the pulse envelope $f(t)$].

\begin{figure}[tb]
\unitlength 1mm
\begin{center}
\begin{picture}(50,35)(0,5)
\put(0,0){\resizebox{50mm}{!}{\includegraphics{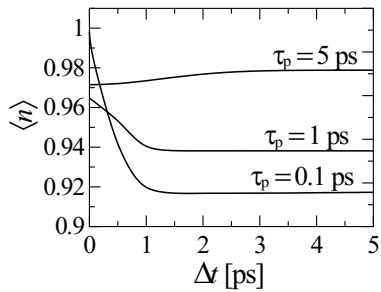}}}
\end{picture}
\end{center}
\caption{\label{fig:2-pulse}The final QD occupation after two
$\pi/2$ pulses separated by time interval $\Delta t$ for pulse
durations as shown.}
\end{figure}

The oscillations are almost perfect for very short pulses ($\sim
1$ ps), then loose their quality for longer pulse durations ($\sim
10$ ps). Although this might be expected from any simple
decoherence model, the striking feature is that the effect
dramatically grows for higher oscillations, despite the fact that
the whole process has exactly constant duration. Even more
surprising is the improvement of the quality of oscillations for
long pulses ($\sim 50$ ps) where, in addition, the first
oscillation is nearly perfect.

For growing number of rotations $n$, the nonlinear pulse spectrum
$S_{\mathrm{a}}(\omega)$ develops a series of maxima of increasing
strength (Fig. \ref{fig:sr}a,b,c). The position of the last and
highest maximum corresponds approximately to $2\pi
n/\tau_{\mathrm{p}}$, in accordance with the semiclassical
resonance concept. However, spectral components are also present
at all the frequencies $2\pi n'/\tau_{\mathrm{p}}$, $n'<n$, which
is due to the turning on/off of the pulse. It is interesting to
note that for high $n$, the low-frequency part of
$S_{\mathrm{a}}(\omega)$ does not evolve with $n$ (Fig.
\ref{fig:sr}d). It is now clear that there are two ways of
minimizing the overlap: either the pulse must be so short that all
the maxima of $S_{\mathrm{a}}(\omega)$ are pushed to the right
into the exponentially vanishing tail of the reservoir spectral
density $R(\omega)$, or the pulse must be very long, to
``squeeze'' the spectral function near $\omega=0$ and thus reduce
its area. In the latter case, the maxima developing with growing
number of oscillations will eventually overlap with $R(\omega)$
destroying the coherence.

Although it might seem that speeding up the process is the
preferred solution, it is clear that this works only because no
high frequency features are included into the present model. In
reality, speeding up the dynamics is limited e.g. by the presence
of excited states and non-adiabatically enhanced LO phonon
coupling (see below). Moreover, it turns out that the resulting
dynamics is actually not fully coherent. It has been shown
\cite{vagov02a} that superposition of states created by an
ultra-short $\pi/2$ pulse becomes corrupted, preventing a second
pulse (after some delay time $\Delta t$) from generating the final
state of $\langle n\rangle=1$ with unit efficiency. In order to
prove the fully coherent character of carrier dynamics it is
necessary to demonstrate the stability of the intermediate state
in a two-pulse experiment. The simulations of such an experiment
are shown in Fig. \ref{fig:2-pulse}. A short pulse
($\tau_{\mathrm{p}}=0.1$ ps) creates a superposition of bare
states (surrounded by non-distorted lattice) which then decohere
due to dressing processes \cite{krummheuer02,jacak03a}. As a
result, the exciton cannot be created by the second pulse with
unit probability \cite{vagov02a}. For a longer pulse
($\tau_{\mathrm{p}}=1$ ps), the lattice partly manages to follow
the evolution of charge distribution during the optical operation
and the destructive effect is smaller. Finally, if the carrier
dynamics is slow compared to the lattice response times
($\tau_{\mathrm{p}}\sim 10$ ps), the lattice distortion follows
adiabatically the changes in the charge distribution and the
superposition created by the first pulse is an eigenstate of the
interacting carrier-lattice system, hence does not undergo any
decoherence and the final effect is the same for any delay time
(its quality limited by decoherence effects during pulsing). In
fact, splitting the $\pi$--pulse into two corresponds to slowing
down the carrier dynamics which, in the absence of decoherence
during delay time, improves the quality of the final state, as
seen in Fig \ref{fig:2-pulse}.

\begin{figure}[tb]
\unitlength 1mm
\begin{center}
\begin{picture}(135,30)(0,13)
\put(0,0){\resizebox{135mm}{!}{\includegraphics{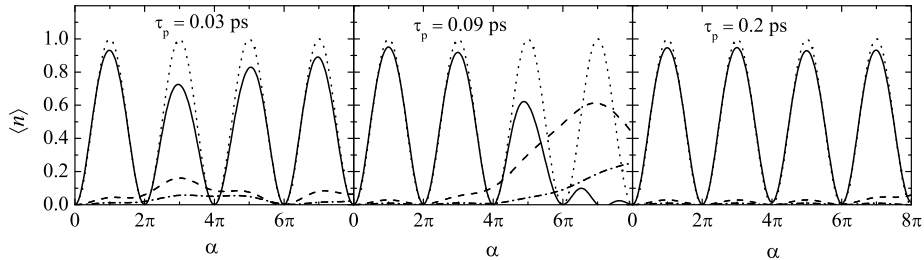}}}
\end{picture}
\end{center}
\caption{\label{fig:opty}Pulse-intensity-dependent Rabi
oscillations for sub-picosecond pulse durations as shown; solid:
ground state, dashed: $M=1, N=0$, dash-dotted: $M=2, N=0$, dotted:
ideal oscillation), where $M$ denotes angular momentum, $N$.}
\end{figure}

\paragraph{Effects of LO phonons.}
The above discussion focused on the relatively long pulse
durations when the contribution from the LO phonons is negligible.
However, an interesting manifestation of the resonance effect may
be observed for sub-picosecond pulses, when the excitation of
longitudinal optical (LO) phonons becomes important (Fig.
\ref{fig:opty}). It should be noted that, although the coupling
between the ground state and the LO modes is strongly reduced by
charge cancelation, effects involving higher (dark) exciton states
may still have considerable impact \cite{jacak03b,fomin98}. First,
let us note that the interaction energy for LO phonons is
comparable with their frequencies which results in a pronounced
reconstruction of the spectrum and appearance of polaron states
\cite{hameau99,hameau02,verzelen02a,jacak03a,jacak02a} which, due
to strong non-diagonal couplings, mix various excitonic levels.
This is manifested in the redistribution of the exciton occupation
among different states even for arbitrarily long and weak pulses,
where the dynamical contribution from the LO phonons vanishes
(Fig. \ref{fig:opty}c). It may be easily verified that in the
limit of slow rotation, when $K_{\mathrm{s,c}}(\epsilon_{l})\to 0$
for $l>1$, the ``static'' occupation of the higher states is equal
to the usual perturbative correction to the state $|1\rangle$.
When $S_{\mathrm{a-c}}(\omega)$ are sufficiently broad, i.e. for
very short pulses or for large number of rotations, the terms
$l>1$ contribute also dynamically. The effect depends on the
relative positions of the narrow LO phonon features and of the
induced dynamics frequencies and the occupation of the excited
states is a non-monotonous function of the pulse duration, due to
the oscillatory character of the spectral functions
$S_{\mathrm{a-c}}(\omega)$ \cite{machnikowski04a}. As seen in Fig.
\ref{fig:opty}, for $\tau_{\mathrm{p}}=0.09$ ps, the frequency of
induced dynamics overlaps with the phonon frequencies, leading to
large damping. For longer durations it shifts towards lower
frequencies, decreasing the impact on the system dynamics.
However, the damping is weaker also for shorter durations
($\tau_{\mathrm{p}}=0.03$, Fig. \ref{fig:opty}a): e.g. for
$\alpha=6\pi$ the largest maximum is now to the right of the
reservoir response frequencies $\omega+\epsilon_{l}$, while for
lower values of $\alpha$ the magnitude of the spectral functions
$S_{\mathrm{a,b,c}}$ is smaller.

\paragraph{Conclusion.}
The above analysis shows that damping of pulse-area-dependent Rabi
oscillations due to interaction with lattice modes is a
fundamental effect of non-Markovian character: it is due to a
semiclassical resonance between the optically induced confined
charge dynamics and the lattice modes. The destructive effect may
be minimized both by speeding up and slowing down the dynamics.
However, in the former case, the system passes through unstable
(decohering) states. Moreover, fast operation on a real system
induces many undesirable effects: transitions to higher states,
bi-exciton generation or resonant LO phonon dynamics. On the other
hand, for slow operation, the number of ``good'' oscillations is
limited. Thus, it is impossible to perform an arbitrary number of
fully coherent Rabi oscillations on an exciton confined in a
quantum dot.

By increasing the pulse duration as
$\tau_{\mathrm{p}}\sim\alpha^{2}$, the phonon effect on the
exciton dynamics may be kept constant. However, in this case the
achievable number of oscillations is strongly restricted by the
exciton lifetime and other (thermally activated) processes.
Eliminating the radiative losses e.g. by using stimulated Raman
adiabatic passage instead of a simple optical excitation
\cite{troiani03} seems to be a promising direction from this point
of view.

The model presented above accounts for the decrease of the quality
of Rabi oscillations in the short duration range observed in the
experiment \cite{borri02a}. It predicts, however, that this trend
is reversed for longer pulse durations. The quantitative value of
96\% for the first maximum of the oscillations with
$\tau_{\mathrm{p}}=1$ ps agrees very well with the experimental
result \cite{zrenner02} although the following extrema are much
worse in reality than predicted here. This suggests an increased
lattice response at higher frequencies which may be due to more
complicated wavefunction geometry \cite{heitz99} or electric-field
induced charge separation leading to strong piezoelectric effects
\cite{krummheuer02}. Nevertheless, as far as it may be inferred
from the experiment \cite{zrenner02}, the decrease of the
oscillation quality seems to saturate after one full Rabi
rotation, as predicted by the model calculations.

\section{Optimal control over a QD qubit}
\label{sec:opty}

\paragraph{Introduction.}
As we have already pointed out, any fast change of the state of
the carrier subsystem leads to spontaneous processes of lattice
relaxation that affect the coherence of the carrier state. In the
previous section we have shown that coherent control is recovered
if the evolution of the carrier subsystem is slow (adiabatic)
compared to the typical timescales of the lattice dynamics. Thus,
the requirement to avoid traces of the carrier dynamics in the
outside world favors slow operation on the carrier subsystem,
contrary to other decoherence processes (of Markovian character),
like radiative decay of the exciton or thermally activated
processes of phonon-assisted transitions to higher states. The
latter have the character of an exponential decay and, for short
times, contribute an error
$\tilde{\delta}=\tau_{\mathrm{g}}/\tau_{\mathrm{d}}$, where
$\tau_{g}$ is the gating time and $\tau_{d}$ is the time constant
of the decay. Means to minimize this contribution by speeding up
the control operation have been proposed by selecting materials to
provide favorable spectrum characteristics \cite{derinaldis02} or
by applying techniques reducing unwanted transitions
\cite{chen01,piermarocchi02}.

In this section we consider the interplay between these two
contribution to the error for the solid-state qubit implementation
using excitonic (charge) states in quantum dots (QDs)
\cite{biolatti00}, with computational states defined by the
absence ($|0\rangle$) or presence ($|1\rangle$) of one exciton in
the ground state of the dot, operated by resonant coupling to
laser light. We show that it leads to a trade-off situation with a
specific gating time corresponding to the minimum decoherence for
a given operation \cite{alicki03}.

\paragraph{The system and its model.}
Since it has been shown experimentally \cite{borri01} that
coherence of superpositions induced by short pulses is unstable,
it seems reasonable to perform operations on \textit{dressed
states}, i.e. on the correctly defined quasiparticles of the
interacting carrier-phonon system \cite{alicki02b}. This may be
formally achieved by employing the solid-state-theory concept of
adiabatic switching on the interaction (as done in
\cite{alicki02a}, cf. \cite{pines89}) to transform the states of
the noninteracting system into the states of the interacting one.
Thus, we assume adiabatic switching on/off of the interaction with
phonons by appending the appropriate exponent to the original
interaction Hamiltonian,
\begin{eqnarray}
H_{\mathrm{int}}=e^{-\varepsilon |t|}\left[ |1\rangle\langle 1|
\sum_{\bm{k}}\left( F_{\bm{k}}b_{\bm{k}}^{\dagger}+
F^{*}_{\bm{k}}b_{\bm{k}} \right) \right], \label{int2}
\end{eqnarray}
where $\varepsilon=0^{+}$. $F_{\bm{k}}$ are the deformation
potential coupling constants between the ground excitonic state
and the longitudinal acoustical phonons (\ref{cpl-X-1}) (this is
the only contributing interaction mechanism for the timescales
discussed here). The operator $S$ now becomes
\begin{displaymath}
    S(t)=U_{\mathrm{C}}(t)e^{-\varepsilon |t|}|1\rangle\!\langle
    1|U_{\mathrm{C}}^{\dag}(t),
\end{displaymath}
where the free evolution operator is obtained from
(\ref{free-evol}) by truncation to the two lowest states. The
general formula (\ref{delta-detal}) may now be used with the bare
state $|\psi_{0}\rangle$. The adiabatic procedure assures that it
is transformed to the dressed state before comparing it to the
density matrix $\rho$, so that the fidelity is defined with
respect to the stable, dressed states.

\paragraph{The irreversible error.}
The form of the operator $Y(\omega)\equiv Y_{11}(\omega)$ is
obtained in the same way as in the previous section. The only
difference with respect to (\ref{Y11}) is that the oscillating
terms now vanish in the long-time limit, due to the adiabatic
switching on/off described above. For the present purpose it is
convenient to write the result in the form
\begin{eqnarray*}
Y(\omega) & = & \frac{1}{\omega}
 F(\omega)(|1\rangle\!\langle 0|-|0\rangle\!\langle 1|
    +|0\rangle\!\langle 0| - |1\rangle\!\langle 1|)  \\
 & & +\frac{1}{\omega}
 F^{*}(-\omega)(|0\rangle\!\langle 1|-|1\rangle\!\langle 0|
    +|0\rangle\!\langle 0| - |1\rangle\!\langle 1|),
\end{eqnarray*}
where
\begin{displaymath}
    F(\omega)=\int_{-\infty}^{\infty}d\tau e^{i\omega\tau}
        \frac{d}{d\tau}e^{i\Phi(\tau)}.
\end{displaymath}

Since in the quantum information processing applications the
initial state of the quantum bit is in general not known, it is
reasonable to consider the error averaged over all input states.
Let us introduce the function
\begin{displaymath}
    S(\omega)=\omega^{2}|\langle\psi_{0}|
        Y(\omega)|\psi_{0}^{\bot}\rangle|^{2}_{\mathrm{av}},
\end{displaymath}
where $|\psi_{0}^{\bot}\rangle$ is a state orthogonal to
$|\psi_{0}\rangle$ and the average is taken over the Bloch sphere.
By restricting (\ref{delta-detal}) to the two-level case, the
average error may now be written
\begin{equation}\label{delta-av}
\delta = \int \frac{d \omega}{\omega^2} R(\omega)S(\omega),
\label{deltaav}
\end{equation}
where $R(\omega)\equiv R_{1}^{(\mathrm{DP})}(\omega)$.

The averaging is most conveniently performed by noting that
\begin{displaymath}
Y(\omega)=\frac{1}{\omega}F(\omega)|+\rangle\!\langle-|
+\frac{1}{\omega}F^{*}(-\omega)|-\rangle\!\langle+|,
\end{displaymath}
where $|\pm\rangle=(|0\rangle\pm |1\rangle)/\sqrt{2}$. Choosing
\begin{displaymath}
\psi_{0} = \cos\frac{\theta}{2}|+\rangle
  +e^{i\varphi}\sin\frac{\theta}{2}|-\rangle,\;\;\;
\psi^{\perp}_{0} = \sin\frac{\theta}{2}|+\rangle
  -e^{i\varphi}\cos\frac{\theta}{2}|-\rangle,
\end{displaymath}
one gets
\begin{displaymath}
S(\omega)=\left|F(\omega)e^{i\varphi}\cos^{2}\frac{\theta}{2}
-F^{*}(-\omega)e^{-i\varphi}\sin^{2}\frac{\theta}{2}\right|^{2},
\end{displaymath}
which, upon averaging over the angles $\theta,\varphi$ on the
Bloch sphere, leads to
\begin{displaymath}
S(\omega)=\frac{1}{12}\left(|F(\omega)|^2+|F(-\omega)|^2\right).
\end{displaymath}

Let us now consider a Gaussian pulse for performing the quantum
gate,
\begin{displaymath}
    f(t)= \frac{\alpha}{\sqrt{2\pi} \tau_{\mathrm{p}}}
    e^{-{1\over 2} ( {t/\tau_{\mathrm{p}} })^2}
\end{displaymath}
Here $\tau_{\mathrm{p}} $ is the gate duration, while $\alpha$ is
the angle determining the gate, e.g. $\alpha={\pi\over 2}$ is the
Hadamard gate, while $\alpha=\pi$ is $\sigma_x$ (bit flip). The
function $|F(\omega)|^2$ that carries all needed information about
spectral properties of the system's dynamics may be approximately
written as
\begin{equation}\label{Fpm}
|F_{\pm}(\omega)|^{2}\approx\alpha^{2} e^{- \tau_{\mathrm{p}}^{2}
\left( \omega\pm \frac{\alpha}{\sqrt{2\pi}\tau_{\mathrm{p}}}
\right)^{2}}.
\end{equation}

As may be seen from (\ref{delta-av}) and (\ref{Fpm}), for a
spectral density $R(\omega)\sim \omega^{n}$ the error scales with
the gate duration as $\tau_{\mathrm{p}}^{-n+1}$ and
$\tau_{\mathrm{p}}^{-n+2}$ at low and high temperatures,
respectively. Therefore, for $n>2$ (typical e.g. for various types
of phonon reservoirs, see section \ref{sec:coupling}) the error
grows for faster gates. Assuming the spectral density of the form
$R(\omega)=R_{\mathrm{DP}}\omega^3$ for low frequencies [in
accordance with (\ref{spdens-DP-onedot}) at $T=0$], we obtain from
(\ref{delta-av}) and (\ref{Fpm})
\begin{displaymath}
    \delta =
    \frac{1}{12} \alpha^2 R_{\mathrm{DP}}
        \tau_{\mathrm{p}}^{-2},\;\;\mbox{at}\;T=0
\end{displaymath}
This leading order formula holds for $\delta \ll 1$. Also, if we
introduce the upper cut-off, the error will be finite even for an
infinitely fast gate (see Fig. \ref{sec:dress}); this is the
ultrafast limit discussed in Section \ref{sec:dress}.

\paragraph{Trade-off between two types of decoherence.}
As we have shown,  in our model the error grows as the speed of
gate increases. This could result in obtaining arbitrarily low
error by choosing suitably low speed of gates. However, if the
system is also subject to other types of noise this becomes
impossible. Indeed, assuming an additional contribution growing
with rate $\gamma_{\mathrm{M}}$, the total error per gate is
\begin{equation}\label{error-simple}
\delta=\frac{\gamma_{\mathrm{nM}}}{\tau_{\mathrm{p}}^{2}}
+\gamma_{\mathrm{M}}\tau_{\mathrm{p}}, \;\;
\gamma_{\mathrm{nM}}=\frac{1}{12}\alpha^{2}R_{\mathrm{DP}},\;\;
\gamma_{\mathrm{M}}=\frac{1}{\tau_{\mathrm{r}}},
\end{equation}
where $\tau_{\mathrm{r}}$ is the characteristic time of Markovian
decoherence (recombination time in the excitonic case). As a
result, the overall error is unavoidable and optimization is
needed. The formulas (\ref{error-simple}) lead to the optimal
values of the form (for $T=0$)
\begin{equation}
\delta_{\mathrm{min}}
=\frac{3}{2} \left(
\frac{2\alpha^{2}R_{\mathrm{DP}}}{3\tau_{\mathrm{r}}^{2}}
\right)^{1/3},\;\; \mbox{for}\; \tau_{\mathrm{p}}
=\left( \frac{2}{3}\alpha^{2}R_{\mathrm{DP}}\tau_{\mathrm{r}}
\right)^{1/3}. \label{eq:minerror}
\end{equation}

\begin{figure}[tb]
\unitlength 1mm
\begin{center}
\begin{picture}(100,28)(0,13)
\put(0,0){\resizebox{100mm}{!}{\includegraphics{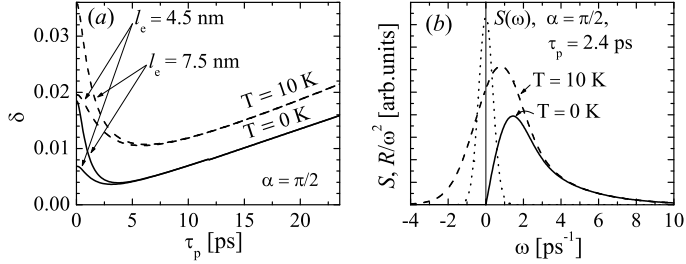}}}
\end{picture}
\end{center}
\caption{\label{fig:result}(a) Combined Markovian and
non-Markovian error for a $\alpha=\pi/2$ rotation on a qubit
implemented as a confined exciton in a InAs/GaAs quantum dot, for
$T=0$ (solid lines) and $T=10$ K (dashed lines), for two dot sizes
(dot height is 20\% of its diameter). The Markovian decoherence
times are inferred from the experimental data
\protect\cite{borri01}. (b) Spectral density of the phonon
reservoir $R(\omega)$ at these two temperatures and the gate
profile $S(\omega)$ for $\alpha=\pi/2$.}
\end{figure}

For the specific material parameters of GaAs, the optimal gate
time and minimal decoherence resulting from Eqs.
(\ref{eq:minerror}) are
\begin{displaymath}
\tau_{\mathrm{p}}=\alpha^{2/3} 1.47\; \mathrm{ps},\;\;
\delta_{\mathrm{min}}=\alpha^{2/3} 0.0035.
\end{displaymath}
The exact solution within the proposed model, taking into account
the cut-off and anisotropy (flat shape) of the dot and allowing
finite temperatures, is shown in Fig. \ref{fig:result}. The
size-dependent cut-off is reflected by a shift of the optimal
parameters for the two dot sizes: larger dots admit faster gates
and lead to lower error.

It should be noted that these optimal times are longer than the
limits imposed by level separation
\cite{biolatti00,chen01,piermarocchi02}. Thus, the non-Markovian
reservoir effects (dressing) seem to be the essential limitation
to the gate speed.

\paragraph{Error reduction by a double-dot encoding.}
As we have seen, for long enough pulses the fidelity of a
one-qubit rotation depends on the low-frequency behavior of the
phonon spectral density. In this section we will show that for a
different encoding of the quantum logical states this behavior can
be made more favorable, even without changing to a different
material. To be specific, we consider a qubit encoded by a single
exciton in a double-dot system, with the logical values
corresponding to the exciton location in one or the other dot. The
physical reason for the weaker lattice impact at low frequencies
is that phonons with wavelengths longer than the distance between
the dots cannot distinguish between the two exciton positions and,
therefore, cannot contribute to decoherence.

As in the preceding subsections, we restrict the discussion to the
two states, denoted $|0\rangle$ and $|1\rangle$, corresponding to
the exciton position in one of the two QDs. We assume that it is
possible to perform rotations in this space, analogous to the
resonantly optically driven rotations of the single-dot excitonic
qubit discussed above. Then, the qubit Hamiltonian
$H_{\mathrm{C}}$ is again the 2-dimensional restriction of
(\ref{ham-rotframe}) with $\Delta_{1}=0$.  The interaction
Hamiltonian is
\begin{equation}
H_{\mathrm{int}} = |0\rangle\langle 0|
\sum_{\bm{k}}\left({F^{(0)}_{\bm{k}}b_{\bm{k}}^{\dagger}
+F_{\bm{k}}^{(0)*}b_{\bm{k}}}\right)\nonumber \\
+|1\rangle\langle 1|
\sum_{\bm{k}}\left({F^{(1)}_{\bm{k}}b_{\bm{k}}^{\dagger}
+F_{\bm{k}}^{(1)*}b_{\bm{k}}}\right),
\end{equation}
where $F^{(0,1)}_{\bm{k}}$ are the coupling constants for the two
exciton localizations. Again, we include only the DP coupling and
suppress the corresponding upper index ``(DP)''. Let us assume
that the dots are placed at $z=\pm D/2$. If we neglect the
possible differences between the geometry of wavefunctions in
these dots then, according to the definitions (\ref{int-DP}) and
(\ref{formfactor}), the coupling constants differ only by a phase
factor,
\begin{displaymath}
    F^{(0,1)}_{\bm{k}}=e^{\pm ik_{z}D/2}F_{\bm{k}},
\end{displaymath}
where $F_{\bm{k}}$ are the coupling constants for an exciton
located at the origin.

With the help of the Weyl operator
\begin{equation}
W = \exp \left[ \sum_{\bm{k}} \left( \gamma_{\bm{k}}b_{\bm{k}}
-\gamma_{\bm{k}}b_{\bm{k}}^{\dagger}\right) \right],\;\;\;
\gamma_{\bm{k}} = \frac{F^{(0)}_{\bm{k}}}{\omega_{\bm{k}}},
\end{equation}
we define the new bosonic operators
\begin{equation}
    \beta_{\bm{k}}=
    Wb_{\mathrm{k}}W^{\dagger}=b_{\mathrm{k}}-\gamma_{\bm{k}}.
\end{equation}
In terms of these, the interaction Hamiltonian has the form (up to
a constant)
\begin{equation}\label{int-2dot}
H_{\mathrm{int}}=e^{-\varepsilon |t|} \left[ |1\rangle\langle 1|
\sum_{\bm{k}}\left({\tilde{F}_{\bm{k}}\beta_{\bm{k}}^{\dagger}
+\tilde{F}_{\bm{k}}^{*}\beta_{\bm{k}}}\right) \right],
\end{equation}
where
\begin{equation}\label{Fk-diff}
\tilde{F}_{\bm{k}} = F_{\bm{k}}^{(1)}-F_{\bm{k}}^{(0)}
=2i\sin\frac{k_{z}D}{2}F_{\bm{k}}.
\end{equation}
The physical meaning of the above manipulation is that the lattice
excitations are now defined with respect to the new equilibrium,
corresponding to the presence of the exciton in one of the dots.
Again, we have introduced the factor describing the
adiabatic switching on/off of the carrier-phonon interaction.

\begin{figure}[tb]
\unitlength 1mm
\begin{center}
\begin{picture}(100,28)(0,13)
\put(0,0){\resizebox{100mm}{!}{\includegraphics{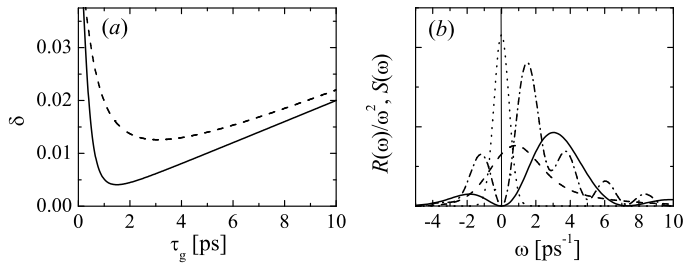}}}
\end{picture}
\end{center}
\caption{\label{fig:2dot} (a) Combined Markovian and non-Markovian
error for a $\alpha=\pi/2$ rotation on a qubit implemented as a
confined exciton in a double InAs/GaAs quantum dot separated by
$D=5$ nm for $T=10$ K (solid line), and for the simple encoding
(dashed line). (b) The spectral densities of the phonon reservoir
at $T=10$ K (solid: $D=5$ nm, dot-dashed: $D=15$ nm, dashed:
single dot) and the nonlinear pulse spectrum (dotted). The
wavefunction width was taken to be $4.5$ nm in-plane and $0.9$ nm
in the growth direction. The difference between electron and hole
localization widths has been neglected.}
\end{figure}

The interaction Hamiltonian is now formally identical to that used
in the previous subsection, but the spectral density now is
\begin{displaymath}
    R(\omega)=R_{\mathrm{DP}}
    \frac{1}{3}\left(\frac{D}{c_{\mathrm{l}}}\right)^{2}\omega^{5}
    [n_{B}(\omega)+1] g(\omega),
\end{displaymath}
where $g(0)=1$. The non-Markovian error for sufficiently long gate
durations at $T=0$ may now be estimated as
\begin{displaymath}
    \delta=\frac{1}{12}\alpha^{2} R_{\mathrm{DP}}
        \frac{1}{3}\left(\frac{D}{c_{\mathrm{l}}}\right)^{2}
        \tau_{\mathrm{p}}^{-4}.
\end{displaymath}
Compared to the corresponding value for the simple encoding
(\ref{error-simple}), this shows much faster decrease with growing
pulse duration. By combining this error with the Markovian
decoherence rate resulting from the final exciton lifetime and
optimizing with respect to the pulse duration one finds
\begin{displaymath}
    \delta_{\mathrm{min}}=\frac{5}{4}\left[
      \frac{2}{9}\alpha^{2}R_{\mathrm{DP}}
       \left( \frac{D}{c_{\mathrm{l}}^{2}}\right)^{2}
       \tau_{\mathrm{r}}^{-4} \right]^{1/5},
       \;\;\;\mbox{for}\;\;
    \tau_{\mathrm{p}}=\left[
        \frac{2}{9}\alpha^{2}R_{\mathrm{DP}}
       \left( \frac{D}{c_{\mathrm{l}}^{2}}\right)^{2}
       \tau_{\mathrm{r}} \right]^{1/5}.
\end{displaymath}
The comparison of the error with single- and double-dot encoding
at $T=10$ K is shown in Fig. \ref{fig:2dot}, with the exciton
decay time the same as in the previous subsection. For the
double-dot encoding, the error is lower by a factor of 3 and the
optimal gating time is shifted to lower values.

\section{Fidelity of a STIRAP qubit}
\label{sec:stirap}

\paragraph{Introduction.}
As we have shown in the preceding sections, the fidelity of
quantum information processing schemes implemented on orbital
(excitonic) degrees of freedom is limited due to the finite
exciton lifetime, usually of order of 1 ps \cite{borri01,bayer02},
and to the adiabaticity requirement excluding unlimited speed-up
of the operation. On the other hand, spin-based proposals
\cite{loss98}, favored by the long lifetime of electron spin
\cite{hanson03} also suffer from serious difficulties: the spin
switching time in typical structures is very long due to weak
magnetic coupling. It seems therefore natural to seek for a scheme
in which the logical values are stored using spin states, while
the operations are performed via optical coupling to the charge
degrees of freedom \cite{imamoglu99,feng03,pazy03a,calarco03}.

Recently, it was proposed \cite{troiani03} to encode the qubit
into spin states of a single excess electron in a QD and perform
an arbitrary rotation \cite{kis02} by employing the stimulated
Raman adiabatic passage (STIRAP) to a state localized spatially in
a neighboring dot \cite{hohenester00c}. Although this passage
requires coupling to a charged exciton ($X^{-}$, or trion) state
which has a finite lifetime, this state is never occupied (in the
ideal case) so that the scheme is not affected by the decoherence
resulting from its decay.

From the discussion presented in the previous sections of this
chapter it is clear that the solid state QD structures, where the
new implementation of these quantum-optical schemes is proposed,
differ essentially from the atomic systems, where these procedures
are successfully applied \cite{bergmann98}. It may be expected
that the coupling to the lattice modes will play an important role
in such sophisticated quantum control schemes,
restricting the range of parameters, where the coherent transfer
may be performed with high fidelity.

In these section we briefly discuss the phonon effects on a STIRAP
qubit. The complete discussion exceeds the scope of this chapter
and may be find in \cite{roszak04a}.

\paragraph{The Raman adiabatic passage.}
The arbitrary rotation of the spin qubit between the states
$|0\rangle$ and $|1\rangle$ (different spin orientations in one
dot) may be performed with the help of an auxiliary state
$|2\rangle$ \cite{kis02} (electron in another dot). All these
three states are coupled to a fourth state (trion) $|3\rangle$ by
laser beams $\Omega_{0},\Omega_{1},\Omega_{2}$. In order to
achieve the Raman coupling, the detunings from the corresponding
dipole transition energies must be the same for all the three
laser frequencies. Therefore, we put
$\omega_{n}=\epsilon_{3}-\epsilon_{n}-\Delta$, $n=0,1,2$. The
envelopes of the first two pulses are proportional to each other,
\begin{displaymath}
    \Omega_{0}(t)=\Omega_{01}(t)\cos\chi,\;\;\;
    \Omega_{1}(t)=\Omega_{01}(t)\sin\chi,\;\;\;\chi\in (0,\frac{\pi}{2}).
\end{displaymath}

Upon transition to the ``rotating'' basis $|\tn
\rangle=e^{-i(\omega_{n}t-\tilde{\delta}_{n})}|n\rangle$,
$\tilde{\delta}_{n}=\delta_{n}-\delta_{0}$, $n=0,1,2$ and
suppressing the constant term, the RWA Hamiltonian may be written
\begin{eqnarray}\label{ham}
    H_{\mathrm{C}} & = & \Delta|3\rangle\langle 3|+
    \frac{1}{2}\Omega_{01}(t) (|B\rangle\langle 3|
    +|3\rangle\langle B|) \\
& & +\frac{1}{2}\Omega_{2}(t)
    \left(e^{i\td_{2}}|\tilde{2}\rangle\langle 3|+
    e^{-i\td_{2}}|3\rangle\langle \tilde{2}|\right). \nonumber
\end{eqnarray}
where
\begin{displaymath}
    |B\rangle =
    \cos\chi|\tilde{0}\rangle+e^{i\td_{1}}\sin\chi|\tilde{1}\rangle,\;\;\;
    |D\rangle =
    -\sin\chi|\tilde{0}\rangle+e^{i\td_{1}}\cos\chi|\tilde{1}\rangle,
\end{displaymath}
Hence, the pulses affect only one linear combination of the qubit
states, the coupled (bright) state $|B\rangle$, while the other
orthogonal combination, $|D\rangle$, remains unaffected. The
Hamiltonian (\ref{ham}) has the eigenstates
\begin{eqnarray}
\label{a0}
  |a_{0}\rangle &=&
    \cos\theta|B\rangle
    -e^{i\td_{2}}\sin\theta|\tilde{2}\rangle, \\
\label{am}
  |a_{-}\rangle &=& \cos\phi
        (\sin\theta|B\rangle
        +e^{i\td_{2}}\cos\theta|\tilde{2}\rangle)
    -\sin\phi|3\rangle,\\
\label{ap}
  |a_{+}\rangle &=& \sin\phi
        (\sin\theta|B\rangle
        +e^{i\td_{2}}\cos\theta|\tilde{2}\rangle)
    +\cos\phi|3\rangle,
\end{eqnarray}
where
\begin{displaymath}
        \tan\theta=\frac{\Omega}{\Omega_{2}},\;\;\;
    \sin\phi=\frac{1}{\sqrt{2}}\left(
    1-\frac{\Delta}{\sqrt{\Delta^{2}+\Omega^{2}+\Omega_{2}^{2}}}
    \right)^{1/2}.
\end{displaymath}
The corresponding eigenvalues are
\begin{equation}\label{trappedstates}
    \lambda_{0}=0,\;\;\;
    \lambda_{\pm}= \frac{1}{2}\left(\Delta \pm
    \sqrt{\Delta^{2}+\Omega^{2}+\Omega_{2}^{2}}\right).
\end{equation}

The system evolution is realized by adiabatic change of the pulse
amplitudes (in this application, the detuning remains constant).
Initially (at the time $s$), both pulses are switched off, hence
$\phi=0$, then $\Omega_{2}$ is switched on first, hence also
$\theta=0$. Therefore, $|a_{0}\rangle$ coincides with $|B\rangle$
and $|a_{-}\rangle$ with $|2\rangle$. During adiabatic evolution
of the parameters, the states move along the corresponding
spectral branches. As shown in Ref. \cite{kis02}, performing the
transfer from $\theta=0$ to $\theta=\pi/2$ and than back with a
different phase $\tilde{\delta}_{2}$ of the $\Omega_{2}$ pulse
results in a rotation in the qubit space $|0\rangle,|1\rangle$
around the axis determined by $\chi$ and by the relative phase
$\tilde{\delta}_{1}$ between $\Omega_{0}$ and $\Omega_{1}$. The
rotation angle is equal to the difference of the
$\tilde{\delta}_{3}$ phases in the first and second pulse
sequence. Ideally, the state $|2\rangle$ is only occupied during
gating, while the state $|3\rangle$ is never occupied.

The above procedure works under assumption that the evolution is
perfectly adiabatic. However, any change of parameters can never
be infinitely slow and the probability of a jump from
$|a_{0}\rangle$ to one of the two other states $|a_{\pm}\rangle$
remains finite, leading to non-vanishing occupation of the trion
state and to decoherence. In order to avoid this error, one has to
impose the usual adiabaticity condition
\begin{equation}\label{adiab-fund}
|\lambda_{\pm}|\gg 1/\tau_{0},
\end{equation}
where $\tau_{0}$ is the duration of the process.

\begin{figure}[tb]
\begin{center}
\unitlength 1mm
\begin{picture}(100,30)(0,13)
\put(0,0){\resizebox{100mm}{!}{\includegraphics{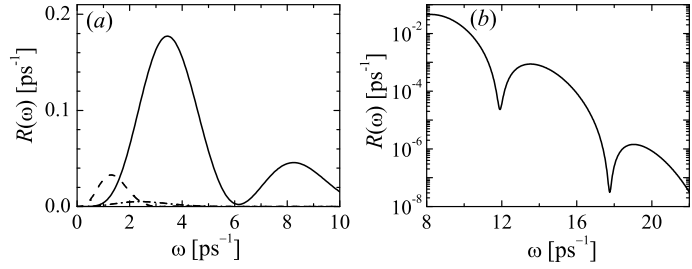}}}
\end{picture}
\end{center}
\caption{\label{fig:Rw-stirap}(a) The contributions to the
spectral density at T=0: DP coupling to LA phonons (solid) and
piezoelectric coupling to TA (dashed) and LA (dash-dotted)
phonons. (b) The high-frequency behavior of the DP contribution.}
\end{figure}

\paragraph{Phonon--induced decoherence.}
In the presence of phonons, the fundamental adiabaticity condition
(\ref{adiab-fund}) is supplemented by the additional requirement
to avoid phonon-assisted processes. It may be shown
\cite{roszak04a} that the only phonon coupling term that
contributes at the leading order of the perturbation is
\begin{equation}\label{H-int-stirap}
    H_{\mathrm{int}}=|\tilde{2}\rangle\langle \tilde{2}|
    \sum_{\bm{k}} \tilde{F}_{\bm{k}}\left(b_{\bm{k}}
    +b_{-\bm{k}}^{\dag} \right),
\end{equation}
where $\tilde{F}_{\bm{k}}$ is obtained in the way analogous to
(\ref{Fk-diff}) by shifting the phonon modes to the equilibrium
appropriate to the occupation of the first dot (states
$|0\rangle,|1\rangle$). This time, however, coupling constants for
a single electron must be used. Vanishing of all the other
contributions results from the indistinguishability of spin states
by phonon interactions, from the large mismatch between the trion
creation energy and phonon energies and from the fact that the
state $|3\rangle$ is not occupied in the ideal case. Since now the
single, uncompensated charge carrier is shifted between different
spatial locations, one may expect a considerable contribution from
the piezoelectric coupling to acoustical phonons. Indeed, as shown
in Fig. \ref{fig:Rw-stirap}, this coupling dominates at low
frequencies, while for high frequencies it decreases rather fast
due to vanishing geometrical factors (\ref{Ml}--\ref{Mt2}) in the
strongest confinement direction. In the high-frequency sector, the
deformation potential coupling dominates, with the oscillatory
tail characteristic to a double-dot structure.

Even though $H_{\mathrm{int}}$ ((\ref{H-int-stirap})) is diagonal
in the bare carrier states, it still gives rise not only to the
pure dephasing effect discussed in the section \ref{sec:opty} but
also to transitions between the trapped carrier-field states
(\ref{a0}--\ref{ap}). The probability of these phonon-induced
transitions becomes very high if the spacing between the trapped
energy levels falls into the area of high phonon spectral density
and the contribution to the error resulting from such transitions
is approximately proportional to the process duration,
\begin{equation}\label{transit}
    \delta\sim R(\lambda_{\pm})\tau_{0},
\end{equation}
with some additional effects appearing due to the pure-dephasing
broadening of the $\lambda_{\pm}$ levels if they are placed in the
narrow local minima in the tail of $R(\omega)$. These strong
decoherence processes may be avoided by either decreasing the
trapped level separation (low-frequency regime, exploiting the
$\omega^{n}$ behavior of spectral densities for $\omega\to 0$) or
increasing it beyond the cut-off (high frequency regime). In both
cases one encounters a trade-off situation, due to the opposite
requirements for phonon-induced jumps (short duration) and for the
fundamental adiabaticity condition and pure dephasing (slow
operation): In the low-frequency regime, avoiding phonon-induced
transitions contradicts the condition for avoiding non-adiabatic
jumps between the trapped states, which may be overcome only by
considerably extending the process duration. In the high-frequency
case, there is competition between the pure dephasing and the
phonon-induced transitions that is overcome by increasing the
trapped state splitting, taking advantage of the particular
structure of the phonon spectral density for a double dot
structure.

\begin{figure}
\begin{center}
\unitlength 1mm
\begin{picture}(100,30)(0,13)
\put(0,0){\resizebox{100mm}{!}{\includegraphics{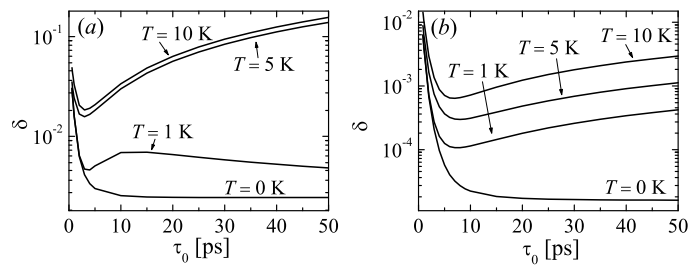}}}
\end{picture}
\end{center}
\caption{\label{fig:opty-stirap}The minimal achievable error as a
function of the transfer duration corresponding to the optimal
pulse parameters with both $|\lambda_{-}|$ in the minimum of
$R(\omega)$ at 6 ps$^{-1}$ (a) and at 12 ps$^{-1}$ (b).}
\end{figure}

In order to provide some quantitative estimations of the error, in
Fig. \ref{fig:opty-stirap} we present the result of the numerical
optimization of the pulse parameters for $|\lambda_{-}|$ located
at the minima of $R(\omega)$ around 6 ps$^{-1}$ and 12 ps$^{-1}$
\cite{roszak04a}. It turns out that a very high fidelity may be
achieved for reasonable pulse parameters and for acceptable
duration of the control sequence (note that $\tau_{0}$ is a
measure of duration of a single transfer; taking into account two
transfers which should be well separated in time leads to the
total duration which is roughly ten times longer).

\section{Conclusions}

We have discussed phonon-related corrections to externally
controlled quantum coherent dynamics of carriers confined in
quantum dots. We have shown that the carrier--phonon interaction
leads to lattice re-lax\-ation (dressing process) af\-ter a change
of the carrier state. This relaxation process leads to
entanglement between the carrier and lattice subsystems and thus
to decoherence of the carrier states.

In order to discuss the possible reduction of this decoherence
effect by slowing down the carrier dynamics, we have developed a
perturbative method for calculating the phonon corrections to
arbitrary evolution. This method allows one to qualitatively
predict the phonon effect based on the nonlinear spectral
properties of the induced carrier dynamics and on the general
knowledge of the spectral properties of the lattice modes.

With this approach, we have studied the Rabi oscillations of
exciton occupation in a QD. We have shown that the lattice
response is resonantly driven by a combination of the (linear)
pulse spectrum and the Rabi frequency. It turns out that the
quality of the oscillations improves both for fast and for slow
dynamics, but only in the later case the intermediate states are
stable.

Applying the perturbative procedure to a QD implementation of a
quantum bit we have shown that the error for a single-qubit
operation may be decreased when one approaches the adiabatic
limit, in which the lattice modes reversibly follow the carrier
evolution. This requires, however, long process durations, which
increases the contribution from other decoherence mechanism. As a
result of the interplay between these two contributions, a
specific pulse duration appears which optimizes the fidelity of
the process. The error may be slightly reduced if the simple qubit
encoding in a single QD is replaced by a double-dot encoding, with
qubit values defined by the exciton position in one of the dots.

Finally, we have discussed the phonon impact on the rotation of a
spin qubit performed optically by adiabatic Raman transition. It
turns out that by a suitable choice of parameters, such a
procedure may be performed with very a high fidelity in a
reasonable time.

The presented results show that phonon-related decoherence
processes form the essential limitation to the coherent
manipulation of confined carrier states in quantum dots. However,
by employing more and more sophisticated control techniques the
resulting error may be substantially reduced. With the help of
theoretical modeling presented in this chapter it is possible to
optimize both the system properties and the operational parameters
in order to maximize the fidelity of quantum coherent control.

\paragraph{Acknowledgments.}
Supported by the Polish Ministry of Scientific Research and
Information Technology under the (so\-licited) Grant No.
PBZ-MIN-008/P03/2003 and by the Polish KBN under Grant No.
PB~2~P03B~085~25. P.M. is grateful to the Humboldt Foundation for
Support.


\end{document}